\documentclass[a4,11pt]{iopart}
\usepackage{graphicx}
\usepackage[usenames]{color}
\usepackage{colordvi}
\usepackage{tikz}

\expandafter\let\csname equation*\endcsname\relax
\expandafter\let\csname endequation*\endcsname\relax
\usepackage{amsmath}
\usepackage{amssymb}

\newcommand{\LL}{L}
\newcommand{\Le}{{\cal L}}
\newcommand{\Ri}{{\cal R}}

\newcommand{\vfi}{v_{\text{fill}}}
\newcommand{\rhostar}{\rho^*}
\newcommand{\sigstar}{\sigma^*}

\definecolor{orange}{rgb}{1.0, 0.5, 0.0}
\definecolor{dg}{rgb}{0.0, 1.0, 1.0}

\def\bluew#1{{}}

\def\sarahcol#1{{\color{black} #1}}

\begin{document}

\title[Spontaneous pulsing states in an active particle system]{Spontaneous pulsing states in an active particle system}


\author{Sarah Klein$^{1,2}$, C\'ecile Appert-Rolland$^{1}$
and Martin R. Evans$^{3}$}

\address{$^1$ Laboratoire de Physique Th\'eorique, Universit\'e Paris-Sud and CNRS UMR 8627 - b\^atiment 210,
91405 Orsay Cedex, France}
\address{$^2$ 
Fachrichtung Theoretische Physik, Universit\"at des Saarlandes, D-66123 Saarbr\"ucken, Germany}
\address{$^3$ SUPA, School of Physics and Astronomy, University of
 Edinburgh,
Peter Guthrie Tait Road,
Edinburgh EH9 3FD, Scotland}

\eads{
  \mailto{mevans@staffmail.ed.ac.uk}, 
}
\vspace*{2ex}
  \date{today}

\today
\begin{abstract}
{
We study a two-lane two-species
exclusion process
inspired by Lin \textit{et al.} (C. Lin \textit{et al}. J. Stat. Mech., 2011),
that exhibits 
a {non-equilibrium} pulsing phase.
Particles move on two parallel one-dimensional tracks, with
one open and one reflecting boundary.
The particle type defines the hopping direction.
When  only particles hopping towards the open end are allowed to
change lane,
the system exhibits a phase transition from a low density {phase} to a pulsing phase
depending on the ratio between particle injection and \bluew{transmutation}{type-changing} rate.
This phase transition can be observed in the stochastic model as well as in a mean-field description. 
In the low density phase, the density profile can be predicted analytically.
The pulsing phase is characterised by a fast filling of the
system and - once filled - by a slowly backwards moving front
separating a decreasing dense region and an expanding low density region. 
The hopping of the front on the discrete lattice is found to create  density
oscillations, both, in time and space.
By means of a stability analysis
we can predict the structure of the dense region during the emptying process,
characterised by exponentially damped perturbations,
both at the open end and near the moving front.
}

\end{abstract}




%

\maketitle



\section{Introduction}
\label{intro}

Driven diffusive systems are studied within physics as 
{models manifesting  non-equilibrium states}. 
They exhibit a richer variety of behaviour and structure of
phases  {than} the traditional equilibrium states~\cite{mukamel2000} such as boundary-induced phase transitions, 
even in one dimension~\cite{krug1991}. Amongst other applications, they have been used to 
model biological processes. For example, the first role of the totally asymmetric exclusion
process with open boundaries (TASEP) was as a model of the dynamics of
ribosomes along mRNA~\cite{macdonald_g_p1968}. Since then the TASEP and its variations have
become the generic starting point for the modelling of intracellular transport effected by molecular
complexes exhibiting steric effects~\cite{chou_m_z2011,aghababaie_m_p1999,appert-rolland_e_s2015}.
Furthermore, at a larger
scale, active media, such as self-phoretic colloids or motile bacteria
may exhibit motility induced phase separated {states~\cite{peruani2012,cates2012,stenhammar2013}.}
At still larger scales biological or artificial birds can exhibit
cohesive flocking states driven by long wavelength sound modes
\cite{toner_t1998,ramaswamy2010}.

Many of these non-equilibrium models have an oscillatory or alternating
nature not exhibited by equilibrium states.  For example the collective direction of a flock 
may switch to and fro~\cite{czirok_b_v1999,oloan_e1999,solon_t2013},
{fish may exhibit intermittency between schooling and milling~\cite{calovi2014},}
proteins may coherently oscillate from end to end of  bacteria~\cite{howard_r_v2001},
and molecular motors may exhibit collective pulsing states~\cite{berg_c2002}.
An understanding of how non-equilibrium conditions allow such
alternating  states is only just emerging and forms an important challenge of non-equilibrium statistical physics.

Returning to the classical case of the TASEP, 
this model has now been generalised to multilane and multispecies versions to incorporate various 
biological features~\cite{chou_m_z2011,appert-rolland_e_s2015}. For
example, in the case of molecular motors,  
the switching of motors between {lanes~\cite{schiffmann_a_s2010,evans2011},}
finite processivity~\cite{parmeggiani_f_f2004}, exchange  
of motors with the cytoplasm~\cite{klumpp_l2003}, and the change of motor states
from active to inactive have been {considered~\cite{nishinari2005a,pinkoviezky_g2013a,pinkoviezky_g2013b,pinkoviezky_g2014b,reichenbach_f_f2006}.}  
Such systems are often only tractable via mean-field (MF) theory, the simplest implementation of which 
results in a set of partial differential equations for the densities of different species on different lanes. 
It has been shown in many systems that the MF equations and furthermore a reduction to an effective 
single lane description gave an accurate description of steady state behaviour in a wide  range of cases~\cite{evans2011,curatolo2016}.
Thus an important task is to determine if and when one can go further with a 
MF description and predict non-equilibrium oscillatory behaviours.

Systems in the half-open tube geometry have been of {particular interest} 
in the  modelling {of} \bluew{cellular} protrusions such as filopodia, lamellapodia, or microvilli
\cite{lan_p2008,wolff2014,pinkoviezky_g2014b}, tubulation~\cite{tailleur_e_k2009} and fungal 
hyphae growth~\cite{sugden2007}. The   boundary condition at the closed end of the tube 
controls the length and may be either growing, treadmilling or no flux.
The model of~\cite{pinkoviezky_g2013a,pinkoviezky_g2013b,pinkoviezky_g2014b}
which incorporates transitions between active and inactive motor states and
treadmilling boundary conditions  (retrograde flow) of 
a fixed length tube, revealed  pulsing behaviour in which bunches of
inactive motors move to the open end of the tube.
In~\cite{lin_s_a2011} a TASEP based model
comprising 13 lanes  and inspired by experimental observations of dynein dynamics on a hyphal tip~\cite{ashwin_l_s2010,schuster2011c} was considered
{(see also  \cite{juhasz}).}
For certain parameter regions the model displays so-called
\textit{pulsing states}  in which
the system alternatively fills and empties, as a result of the motion
of some fronts separating low- and high-density regions.

We will study here a simplified\footnote{Actually the model considered here
is a special case of the very general model introduced at the beginning of~\cite{lin_s_a2011}.} two-lane version of the latter model
\cite{lin_s_a2011} that also exhibits these pulsing states 
but is more amenable to analysis.
Our aim is to identify  the minimal ingredients required to give
pulsing states and to gain some theoretical  understanding through
analysis of a MF description.

The paper is organised as follows.  In section 2 we introduce a 
two-lane two species stochastic model and present numerical
simulations which illustrate that two phases of the model---low
density and pulsing---exist. In section 3 we study the MF
approximation to the model and show that the same two phases also
exist.
In the remainder of the paper we choose to concentrate
on the MF version of the model.
We then proceed in section 4 to analyse first the low density phase.
within the MF approximation.
In section 5 and 6 we turn to the
pulsing phase and analyse the filling and emptying phases of the
oscillatory behaviour respectively. 
The analysis reveals some subtle aspects of the oscillatory
state and interestingly we will show that the discreteness of the
lattice plays an unexpected role in the dynamics of the system.
We then  carry out a
stability analysis of flat profiles in the high density region in
section 7 and we examine the transition between the low density phase
and the pulsing phase in section 8. Finally we conclude with a
discussion in section 9.



\section{Stochastic Particle Model}
\label{sec:particle}

\subsection{Definition of the model}
\begin{figure}[tb]
\includegraphics[width = \textwidth]{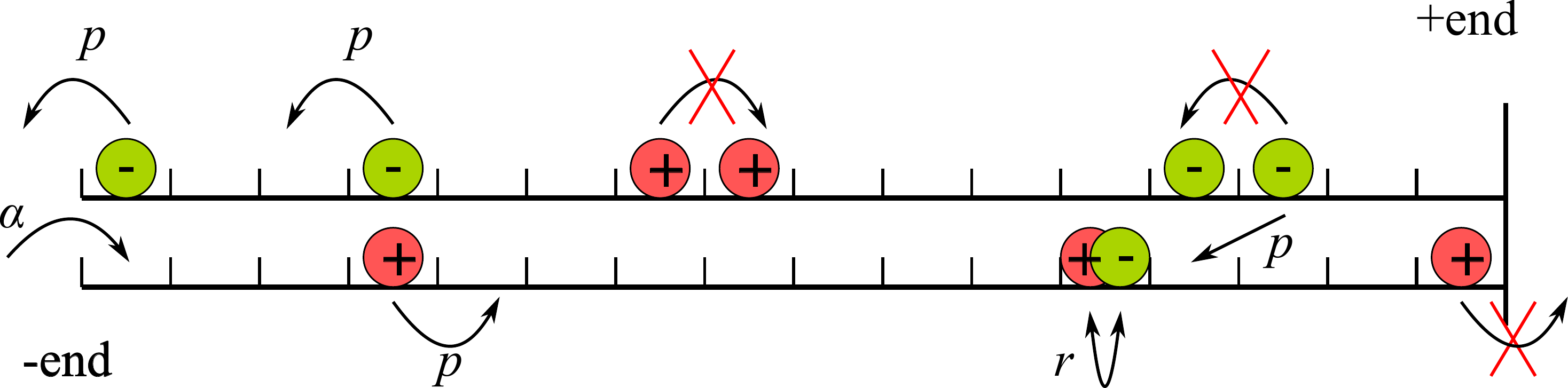}
\caption{Schematic drawing of the possible transitions in the two lane model.
Red (plus) particles \bluew{can} hop with rate $p$ from left to right.
Green (minus) particles hop with rate $p$ from right to left.
Minus particles \bluew{can} change lane with rate $p$ if the next site
on their lane is occupied while it is empty on the other lane.
Particles transmute with rate $r$.
Injection of plus particles occurs with rate $\alpha$ on both entrance sites
on the left, if they are empty.
Minus particles located on the leftmost sites leave the system
with rate $p$.}\label{skizze_2l}
\end{figure}

In this section we define a stochastic particle model, inspired by Lin \textit{et
al}. \cite{lin_s_a2011}, which exhibits pulsing behaviour  in
a certain parameter range.
We consider a lattice composed of
two parallel lanes of
length $\LL$ 
with an open boundary on the left and a closed boundary on the
right (see Figure \ref{skizze_2l}).
Two types of point-like particles hop stochastically on the lattice.
Plus particles are injected at the left end of each lane
with rate $\alpha$,
and hop from site to site to the right with rate $p$.
Minus particles hop only towards the left, also with rate $p$,
and leave the system with rate $p$ at the left boundary.
Additionally, particles {change type}  (and consequently reverse
their hopping direction) with rate $r$. We refer to the type-changing
process as transmutation.
The particle position is exclusive so that a single
lattice site can only be occupied by exactly
one particle or else be vacant. 

In addition to these dynamical processes,
a minus particle can change lane if the next site (to the left)
on the same lane is occupied,
and the next site on the other lane is empty.
This transition occurs with the same 
rate $p$ as the hopping within one lane. The plus particles on the
other hand do not
change lanes.
Thus, the lane changes of minus particles  couple the
dynamics of the two lanes. 

If the ratio
 $p/r$ of hopping rate to transmutation rate  is too small, {plus} particles will {transmute and} turn back towards
the left end before having travelled through the whole system.
In this paper, we  consider only the case where
$p/r$ is at least of order $\mathcal{O}(\LL) $, in order
to allow particles to reach the reflecting boundary after they have entered at
the left.

{
If the ratio $\alpha/p$ is too large, 
{minus particles are impeded from leaving the left-hand end}
and the dynamics slows down. To keep simulations
within reasonable time, we restricted ourselves
to the case $\alpha/p \ll 1$, i.e. at most of order $\mathcal{O}(1/\LL)$.
}

\subsection{Phases of the system}
\label{sec_phases_PM}
\begin{figure}[h!]
\begin{minipage}{0.5\textwidth}
\centering\includegraphics[width = \textwidth]{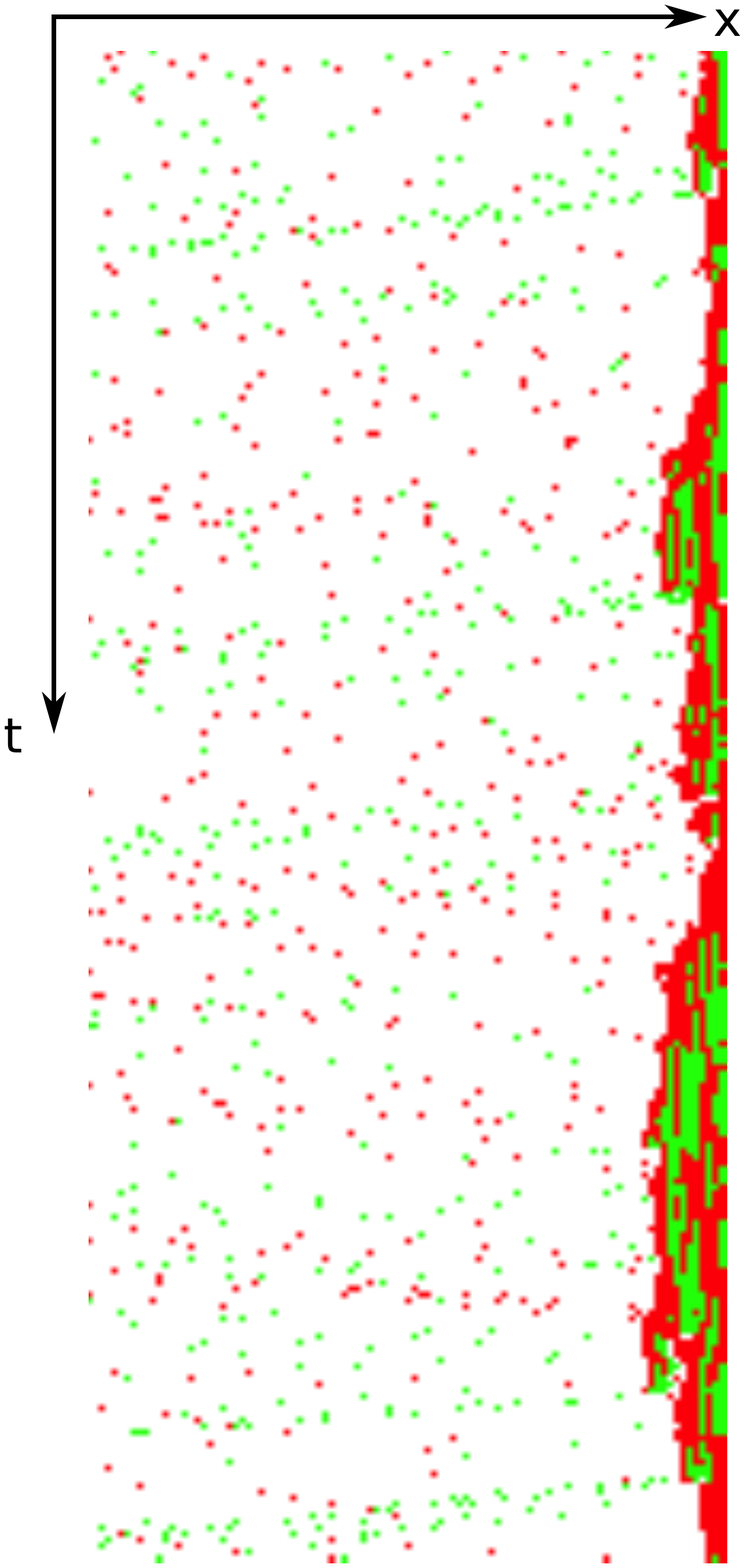}
(a)
\end{minipage}
\begin{minipage}{0.5\textwidth}
\centering\includegraphics[width = \textwidth]{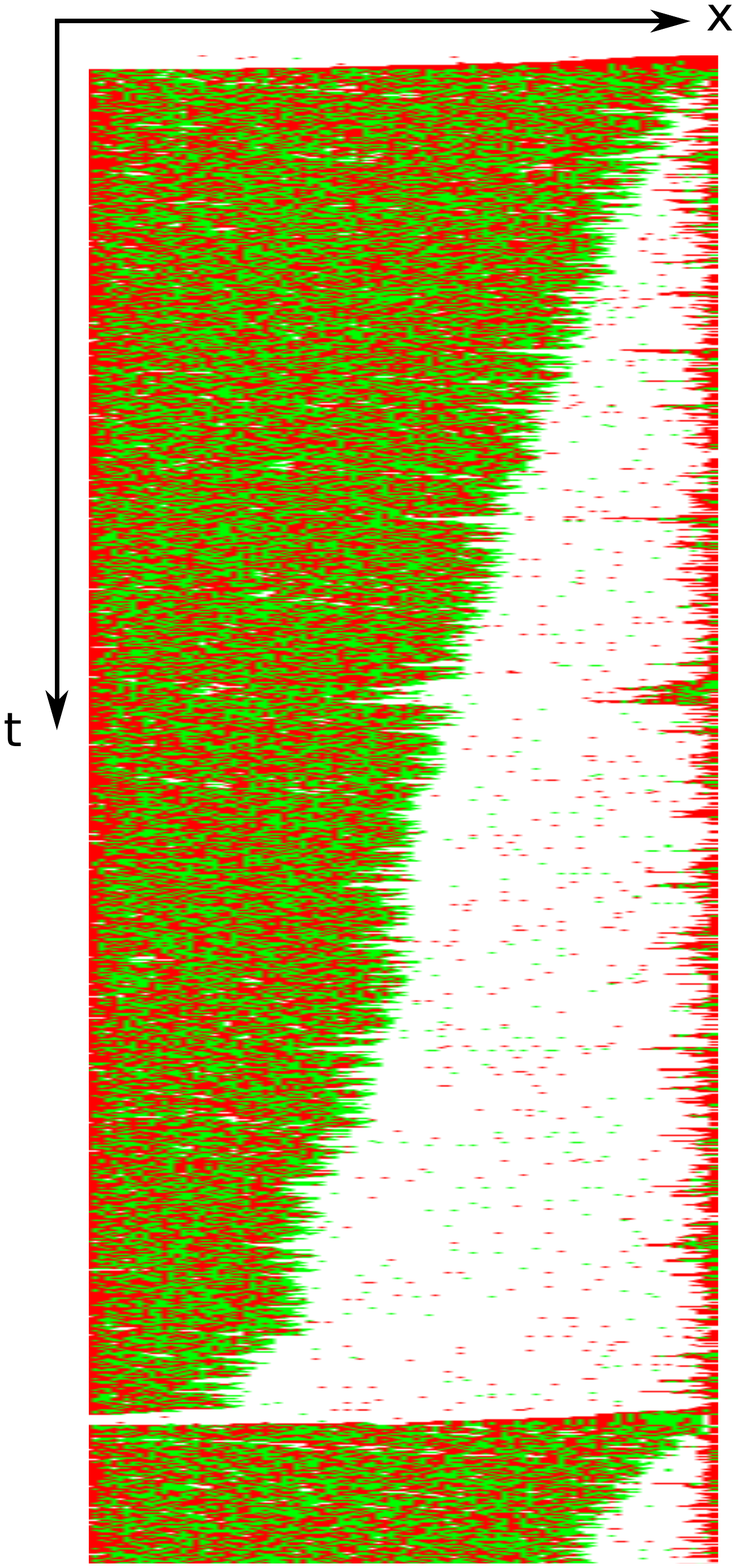}
(b)
\end{minipage}
\caption{
(a-b) Spatio-temporal plot of one lane in the two-lane Particle Model,
in the (a) low density phase, and in the (b) pulsing phase.
Red (plus) particles move to the right, green (minus) particles
to the left. 
Parameters: $\LL=100$, $p=200$ for both, (a) $r=1$ and $\alpha = 5$, (b) $r = 0.04$ and $\alpha=4$.
}\label{kymo}
\end{figure}

For $\alpha$ sufficiently small 
the system is in a {\em low density phase}.
This phase is characterised by a low density of particles in the bulk
of the system, 
possibly with a boundary layer of accumulated plus particles
at the reflecting boundary {at the right-hand end} (see Figure \ref{kymo} (a)).

With increasing $\alpha$ a phase transition from a stationary state with 
zero flux to a \textit{pulsing state} {occurs}.
This state is characterised by a 
cyclic behaviour. First the system rapidly fills up with plus particles. 
Once the total density throughout the system is close to $1$,
an emptying process begins and
a front separating a high density and a low density region moves 
from the reflecting boundary
towards the open one
(see the spatio-temporal plot in Figure \ref{kymo} b). 
When the system is almost empty, the filling process starts again.
These cyclic density changes result in a change of the sign of the current, as the system empties and fills 
and so particles flow in and out. 
{Just as} in the low density phase, a small accumulation of plus particles
persists near the wall on the right. The filling and emptying process
continues indefinitely with a regular period which depends on $r$.

Note that the pulsing phase can exist only if there is an asymmetry in the dynamics of
plus and minus particles \cite{lin_s_a2011}.
The role of this asymmetry is to induce in dense regions a net particle flow
towards the exit that will be responsible for the emptying of the system.
This net flow can be obtained either by allowing only minus particles
to change lane --- a feature that will make them more motile than plus particles 
in dense regions ---
or by producing more minus particles than plus particles through asymmetric transmutation rates, 
as was considered in~\cite{lin_s_a2011}.
In the latter case pulsing states can be observed even when both types of particles
can change lanes.
In the present paper, we  only consider the case of asymmetric
lane changing rules.

A third trivial phase is observed for $r=0$: 
starting from an empty lattice plus particles  
fill the whole system and can never exit the system.

\clearpage

\section{The Mean-Field model}

{In this section} we introduce a mean-field approximation
to the stochastic particle model of the previous section and
we will show that it exhibits the same phases and phenomenology,
in particular,  the same pulsing behaviour.
Importantly, the MF equations are amenable to analysis
and allow us a deeper insight into the mechanisms
involved. Consequently, the MF model in its own right becomes the subject of our  study.

\subsection{Definition}

We now write down the equations for the time evolution
of the densities in the MF approximation where 
all correlations between densities are ignored. To this end
we replace the boolean occupation numbers at site $i$ by
the densities $\rho_i$ and $\sigma_i$,
for plus and minus particles respectively.
In the MF model we assume not only that density correlations are
factorised, but also that the densities are the same on both lanes
for a given location $i$ (this latter assumption is implicit in
the fact that $\rho_i$ and $\sigma_i$ depend only on $i$ but not
on the lane). In this way we arrive at the following system of 
bulk and boundary equations. 

\noindent Bulk equations:
\begin{subequations}
\begin{eqnarray}
\frac{\text{d}\rho_i}{\text{d}t} = && 
p \rho_{i-1} (1-\rho_{i} - \sigma_{i})
- p \rho_i (1-\rho_{i+1} - \sigma_{i+1})
 \nonumber \\
\label{bulk_2l_p1}
&&- r(\rho_i -\sigma_i)  \\ \nonumber \\ \nonumber
\frac{\text{d}\sigma_i}{\text{d}t} =  &&
p \sigma_{i+1} \left[1-(\rho_{i} + \sigma_{i})^2\right]
- p \sigma_i \left[1-(\rho_{i-1} + \sigma_{i-1})^2\right]
 \\ 
\label{bulk_2l_s1}
&&+ r(\rho_i - \sigma_i) .
\end{eqnarray}
\end{subequations}

\noindent Left boundary:
\begin{subequations}
\begin{eqnarray}
\label{left_2l_1}
\frac{\text{d}\rho_1}{\text{d}t} & = & \alpha (1-\rho_{1} - \sigma_{1}) - p \rho_{1} (1-\rho_{2} - \sigma_{2}) 
 - r(\rho_1 - \sigma_1) \\ \nonumber \\
\label{left_2l_2}
\frac{\text{d}\sigma_1}{\text{d}t} & = &
 p \sigma_{2} \left[1-(\rho_{1} + \sigma_{1})^2\right]
 - p \sigma_1
 + r(\rho_1 - \sigma_1).
\end{eqnarray}
\end{subequations}

\noindent Right boundary:
\begin{subequations}
\begin{eqnarray}
\frac{\text{d}\rho_\LL}{\text{d}t} & = & 
 p \rho_{\LL-1} (1-\rho_{\LL} - \sigma_{\LL}) 
  - r(\rho_\LL - \sigma_\LL) \label{right_2l_1}
\\ \nonumber \\
\frac{\text{d}\sigma_\LL}{\text{d}t} & = & 
- p \sigma_\LL \left[1-(\rho_{\LL-1} + \sigma_{\LL-1})^2\right]
+ r(\rho_\LL - \sigma_\LL) .
\label{right_2l_2}
\end{eqnarray}
\end{subequations}

In equations (\ref{bulk_2l_p1}, \ref{bulk_2l_s1}), 
one recognises the bulk fluxes of plus/minus particles
through the link $(i,i+1)$,
which we  denote 
by
\begin{subequations} \label{fluxdef}
\begin{align}
\label{fluxdef:pl}
&J^+_i = 
 p \rho_i (1-\rho_{i+1} - \sigma_{i+1}) \\
\label{fluxdef:mi}
&J^-_i = - p \sigma_{i+1} [1-(\rho_{i}+\sigma_{i}) ^2].
\end{align}
\end{subequations}
Note that, as fluxes are measured to the right,
the flux of plus particles gives a positive
value for $J^+_i$ whereas
the flux of minus particles gives a negative one
for $J^-_i$ so that $J^+_i$ and $J^-_i$ are always of opposite sign.

The boundary conditions coming from  
(\ref{left_2l_1}--\ref{right_2l_2})
can  be rewritten
\begin{subequations} \label{fluxdefbc}
\begin{align}
\label{fluxdefbcl:pl}
&J^+_0 = 
 \alpha (1-\rho_{1} - \sigma_{1}) \\
\label{fluxdefbcl:mi}
&J^-_0 = - p \sigma_{1} \\
\label{fluxdefbcr}
&J^+_\LL = J^-_\LL = 0.
\end{align}
\end{subequations}

Equations (\ref{bulk_2l_p1}-\ref{right_2l_2})
can be considered as defining a dynamical system in its own right,
which we  refer to as the MF model in the remainder of this paper.
This set of equations
can directly be solved numerically.

\begin{figure}[htb]
\begin{minipage}{0.45\textwidth}
\centering\includegraphics[width=0.9\textwidth]{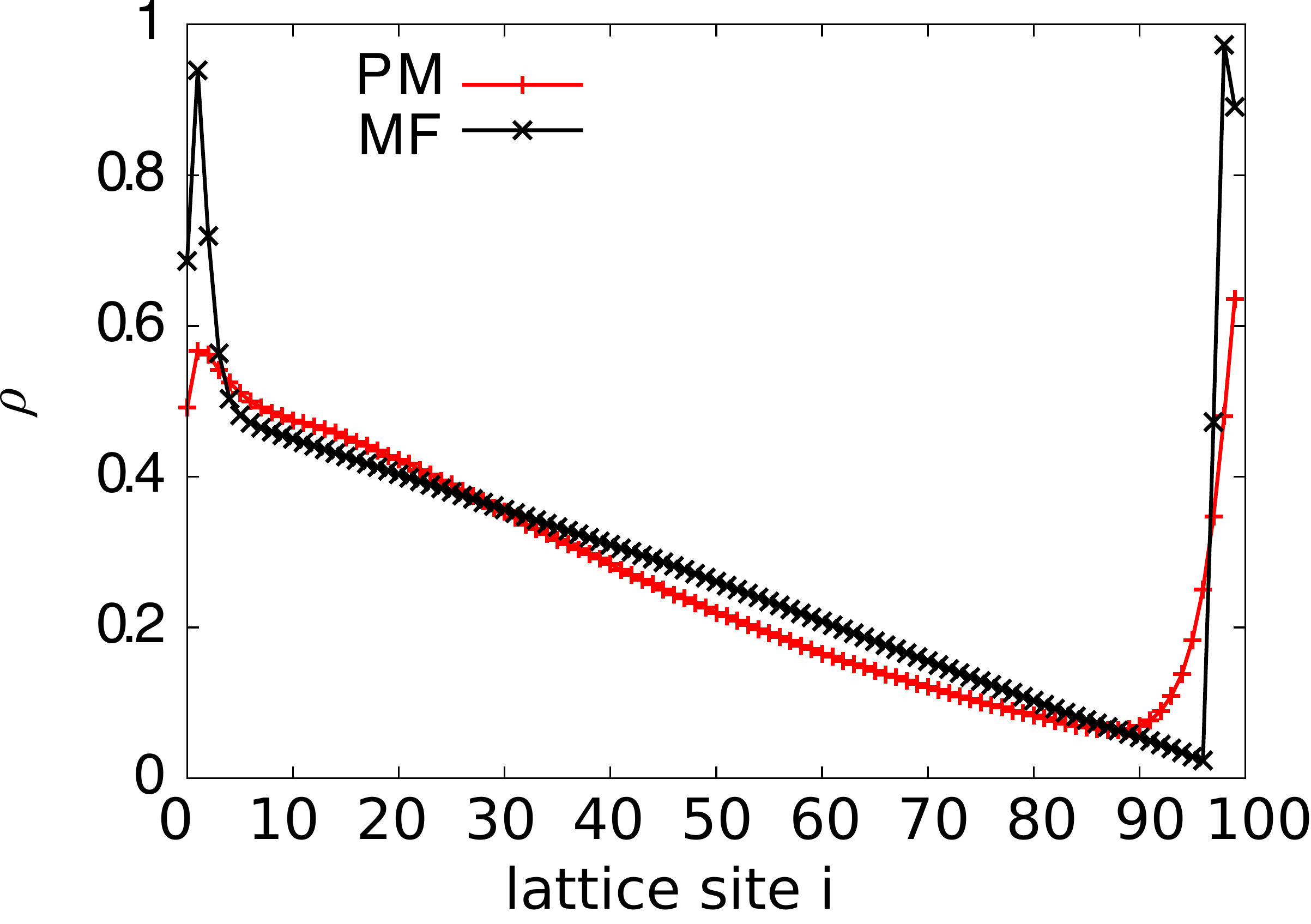}\\
\centering(a)
\end{minipage}
\begin{minipage}{0.45\textwidth}
\centering\includegraphics[width=0.9\textwidth]{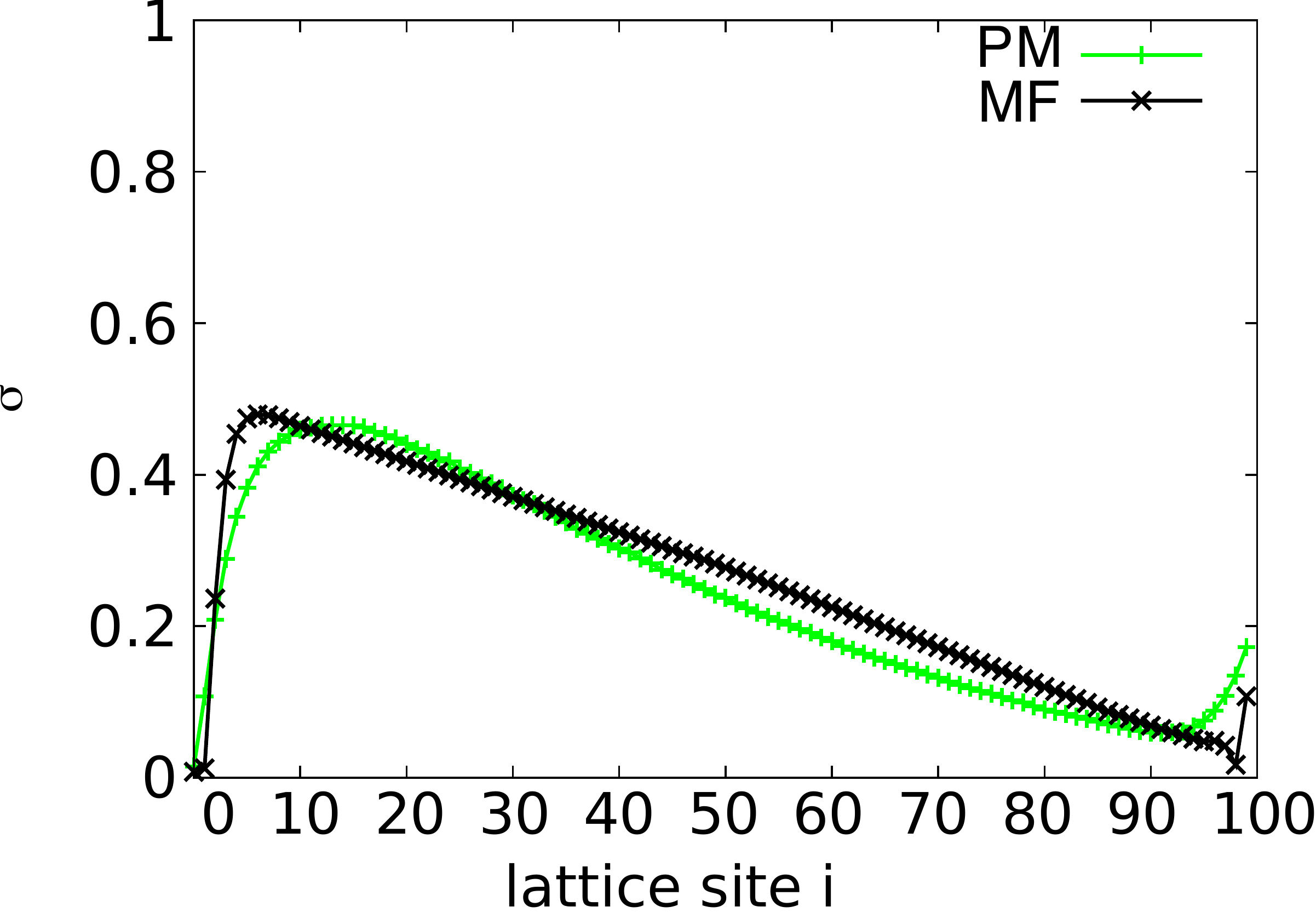}\\
\centering(b)
\end{minipage}
\hspace*{2cm}\centering\includegraphics[width=0.35\textwidth]{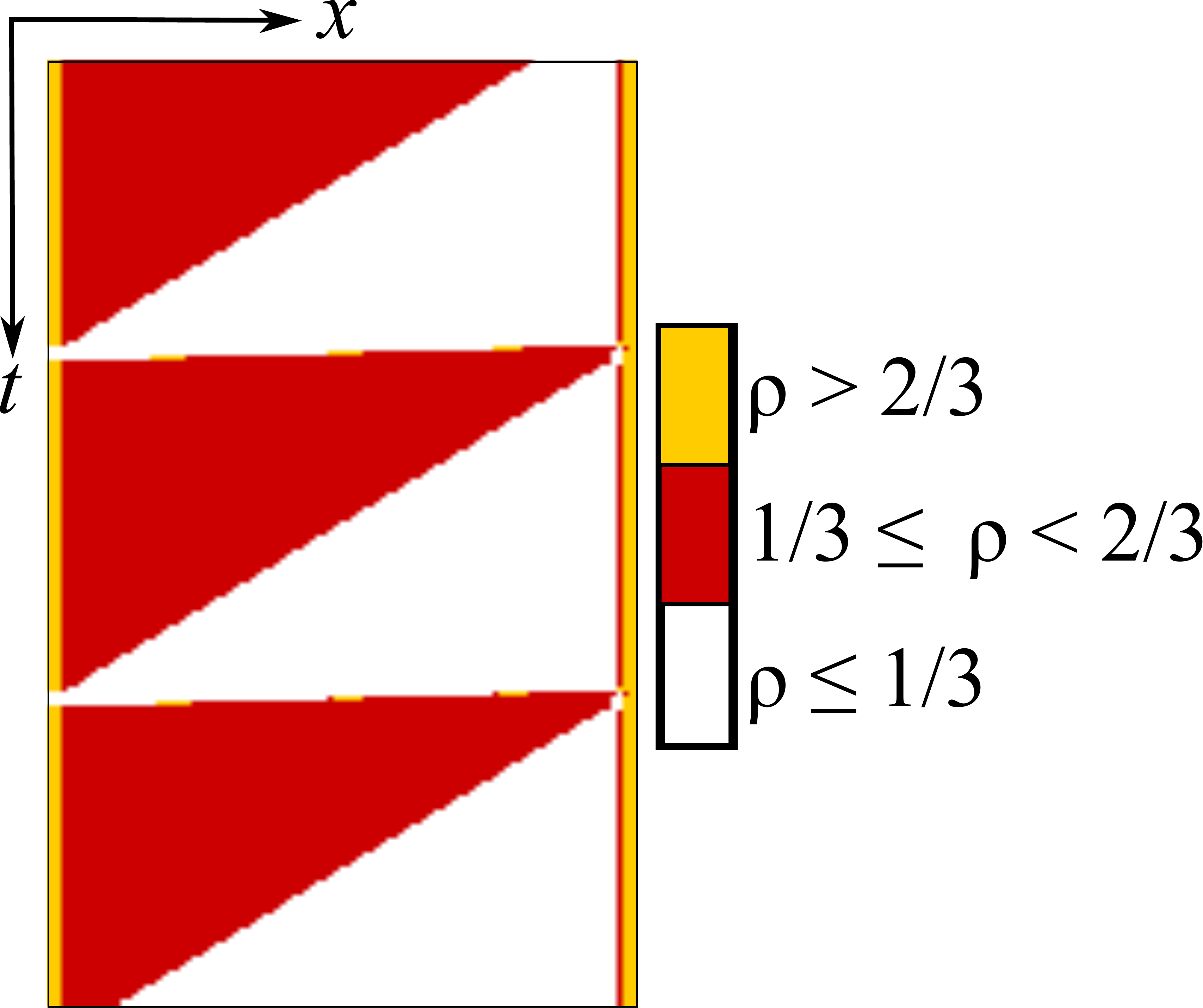}\\
\hspace*{2cm}\centering(c)
\caption{Time-averaged density profiles for $\rho$ (a) or $\sigma$ (b)
obtained from Monte Carlo simulations of the two lane particle model (PM-red/-green)
and from the MF equations (MF-black). In (c) a spatio-temporal plot of the MF model is shown. \\
Parameters: $p=200$, $\alpha=5$, $r=0.5$ and $L=100$.
}\label{density_2l}
\end{figure}

The pulsing states which are found in the Particle Model 
of section~\ref{sec:particle} are also observed in the MF 
model, as shown in Figure \ref{density_2l}.
The Particle Model and the MF model are not completely equivalent,
as can be seen for example from the slight
differences in the time-averaged density profiles in Figure \ref{density_2l} (a),(b).
The comparison of Figure~\ref{kymo}(b) and \ref{density_2l}(c) shows
also the role of fluctuations at the end of the emptying process
in the Particle Model.

First we give an overview of the various phases
exhibited by the system in the MF description, before studying them in more detail in later sections.
All figures or analytical calculations presented
in the remainder of this paper will refer to the MF model.

\subsection{Phase diagram}

\begin{figure}[htb]
\centering\includegraphics[width = 0.6\textwidth]{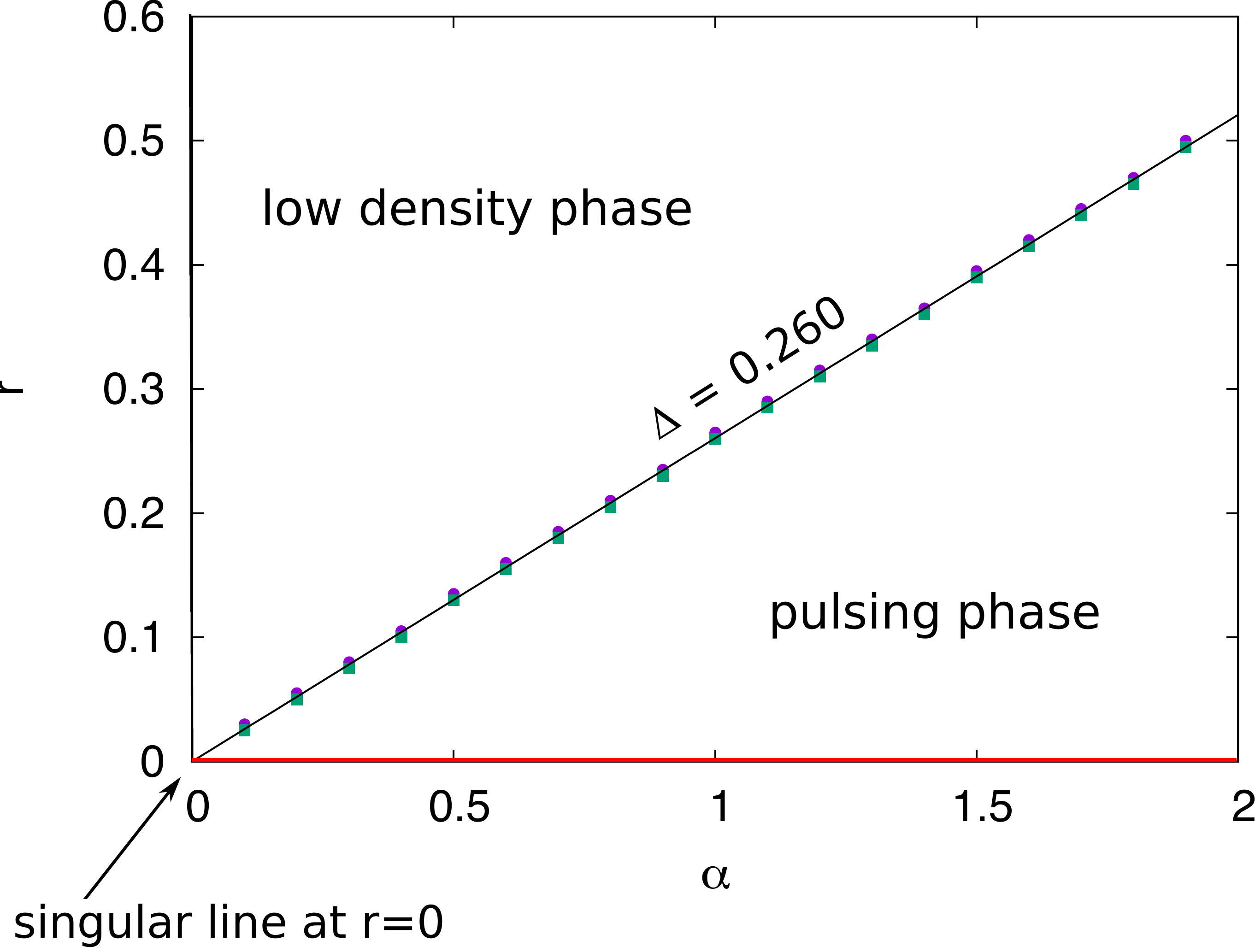}
\caption{Phase diagram for the MF model with $p=200$ {and $L=100$}.
The system is for
$r>0$ either in a pulsing or a low density phase. For $r=0$ the system fills up
with plus particles and remains blocked in a state of density one (red line).
Purple circles (resp. green square) belong to the low (resp. pulsing) phase.
The straight line is a guide to the eyes that separates the different types
of symbols and has slope $\Delta=0.260$.
{
We checked for $p=400$, $600$ and $1000$ that this slope
does not depend on $p$. For $p=600$, we found no dependence
of the slope for $L=100$ and $L=300$.
}
}\label{phase_diagram}
\end{figure}

\begin{figure}[htb]
\centering\includegraphics[width = 0.7\textwidth]{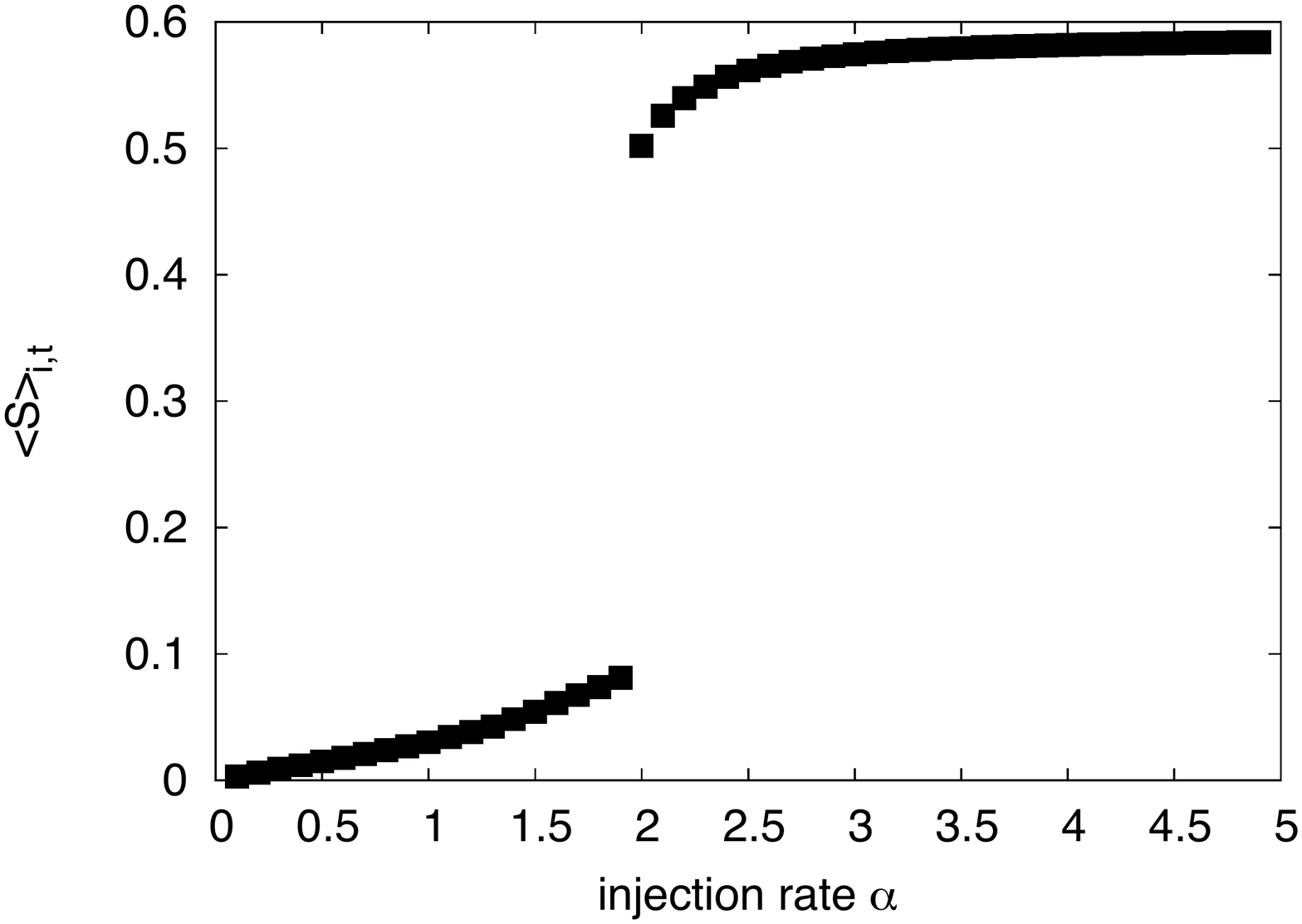}
\caption{Average density per site $\langle S \rangle_{i,t}$,
averaged over the whole system and over time, in the MF model.
Here $r = 0.5$, $p=200$, $L=100$.
}\label{phase_transition}
\end{figure}


We consider the MF model
only in the regime where $p/r$ is at least of order $\mathcal{O}(\LL)$
(as for the particle model).
In the MF model we find the same three  phases identified in Section~\ref{sec:particle}~: a low density phase, a pulsing phase, and a singular 
line
for $r=0$, where the system fills with plus particles and remains blocked
in a state with density one.
In the remainder of this paper  we only consider the non-trivial case where $r>0$.
In Figure \ref{phase_diagram} we see that the phase boundary between
the low density phase at high $r$/low $\alpha$ 
and the pulsing phase at  low $r$/high $\alpha$ 
appears to be a straight line in the $r$--$\alpha$ plane.
{
For the explored parameter values (see Figure~\ref{phase_diagram}), we did not
find any dependence of the slope on $p$ or $L$.
}
Thus the system's phase is determined by the ratio
between $\alpha$ and $r$,
at least when they are small compared to $p$. 

As an order parameter for the system
{we  take the total density per site,
\begin{equation}
\quad S_i \equiv \rho_i+\sigma_i\;,
\end{equation}
averaged over the whole system and over time,
which we denote $\langle S \rangle_{i,t}$.}
Then as illustrated
in Fig~\ref{phase_transition}, the order parameter exhibits a discontinuity at the transition.


\begin{figure}[htb]
\centering\includegraphics[width = 0.7\textwidth]{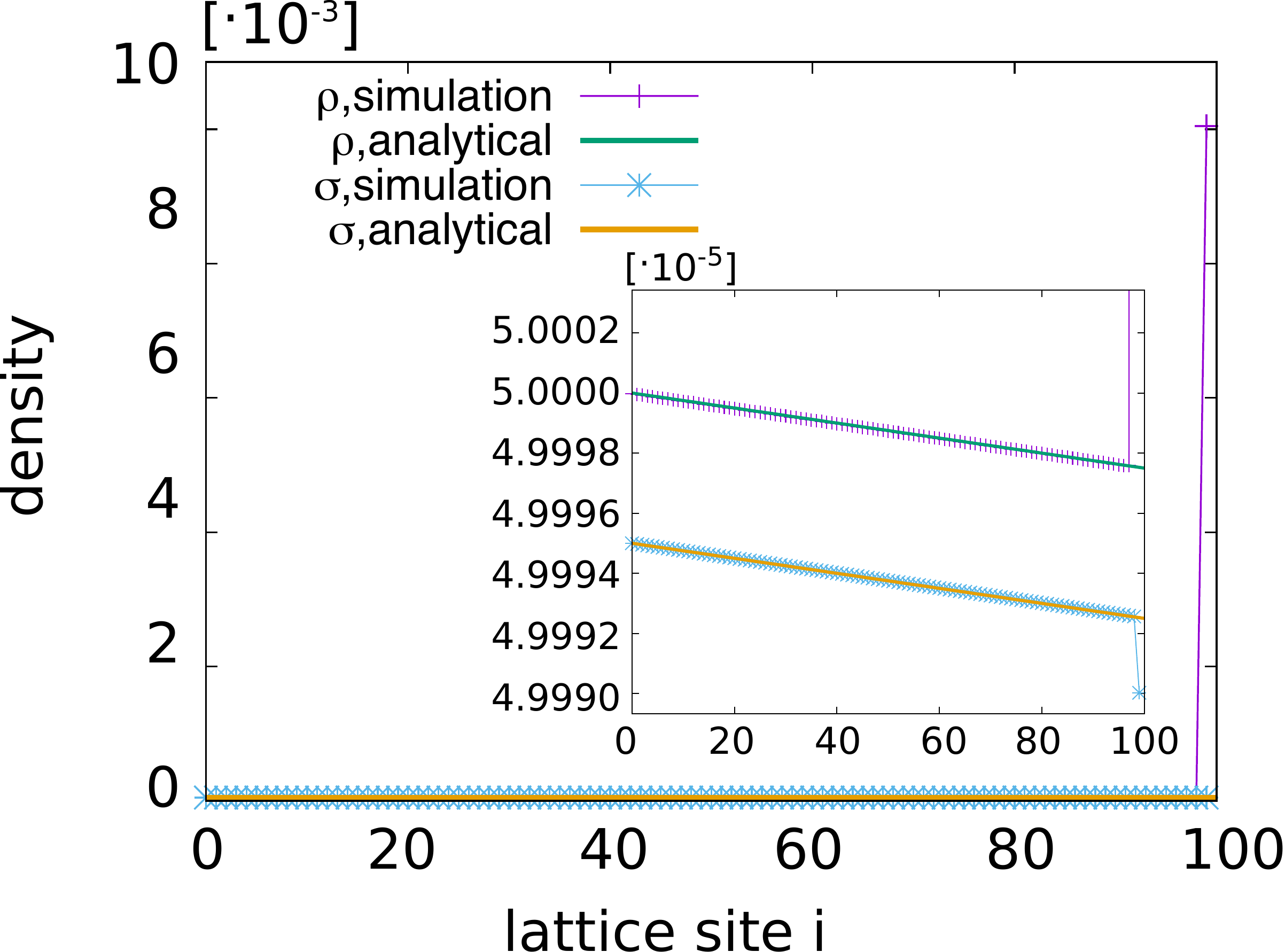}
\caption{Density profile in the low density phase, for $\alpha=2/\LL$, $r=1$, and $p=2\LL$ with $\LL=100$. The inset shows the good agreement between the
prediction (\ref{rho_lowdens},\ref{sigma_lowdens}) and the numerical results.
}\label{free_flow_dens}
\end{figure}

\subsection{Total Current}

It is useful to consider
the net flux of particles or total current $K_i$ through the link $(i,i+1)$
which is given by
\begin{align}
K_i & \equiv  J^+_i + J^-_i\;.
\end{align}
As a consequence of mass conservation, we have the continuity equation
\begin{equation}
\label{bulk_mass}
\frac{\text{d} S_i}{\text{d}t} = 
K_{i-1} - K_i 
\end{equation}

In a  stationary state,
$K_i$ is constant with respect to $i$
and equal to zero because of the  reflecting boundary of the system:
\begin{equation}
K_i^{*} = 0
\label{Knul}
\end{equation}
where the superscript $\mbox{}^{*}$ indicates that the variable is taken in the stationary state.
Therefore, in a  stationary state,
at the entrance
the net flux must also be zero
\begin{align}
K_0^{*} = 0 = \alpha (1-S_1^{*}) - p \sigma_1^{*}\;.
\end{align}
This yields the relation
\begin{equation}
\sigma_1^{*} = \frac{\alpha}{\alpha+p}(1-\rho_1^{*}) \le \frac{\alpha}{\alpha+p} < \frac{\alpha}{p}\;.
\label{s1_ness}
\end{equation}
As we always consider that $\alpha/p \ll 1$,
 one finds from (\ref{s1_ness}) that in a stationary state
\begin{equation}
\sigma_1^{*}
\ll 1.
\end{equation}
Note that this is true only for $\sigma_1^{*}$, while
$\rho_1^{*}$ can be either small or large (large meaning here
close to $1$), depending on the value of $r$.

\section{Analysis of the low density phase}
\label{sec:lowdens}

In this section we assume not only
that $\alpha/p \ll 1$,
but also that we are in the low density phase,
so that in the stationary state, not only $\sigma_1^{*}$
but also $\rho_1^{*}$ is small.
To fix ideas, we assume in the remainder of this section that
\begin{equation}
p = \mathcal{O}\left(\LL\right), \;\;\;
r = \mathcal{O}\left(1\right),  \;\;\;\mbox{ and }
\;\;\alpha = \mathcal{O}\left(\frac{1}{\LL}\right).
\label{eq_scaling}
\end{equation}

Assuming smooth variations of the densities in the system,
one can derive the partial differential equations describing
the behaviour of the system in the continuous space limit.
This description is indeed expected to be relevant in the low density
limit for which there are no sharp shocks (except near
the reflecting boundary).


We introduce a continuous space coordinate $x=i/\LL$ and
assume that $\LL$ is large. Then, setting $\rho_i(t) \to \rho(x,t)$
and $\sigma_i(t) \to \sigma(x,t)$, the Taylor expansion
of Eqs~(\ref{bulk_2l_p1}, \ref{bulk_2l_s1})  gives
\begin{subequations}
\begin{eqnarray}
\frac{\partial \rho}{\partial t} &=& 
- r \left( \rho - \sigma\right) 
+ \frac{p}{\LL} \frac{\partial}{\partial x}
\left[\rho(\rho+\sigma-1)\right]+\cdots
%
\label{taylor_exp_rho} \\
\frac{\partial \sigma}{\partial t} &=& 
r \left( \rho - \sigma\right)
+ \frac{p}{\LL} \frac{\partial}{\partial x}
\left[\sigma\left(1-(\rho+\sigma)^2\right) \right] +\cdots
\label{taylor_exp_sig}
\end{eqnarray}
\end{subequations}
{It should be noted that corresponding equations linearised in the densities
have been considered in \cite{lin_s_a2011} appendix A. }

We are interested in the stationary state of (\ref{taylor_exp_rho}, \ref{taylor_exp_sig}),
so that the left-hand side of the equations 
is zero.
For simplicity, we will omit the superscript $\mbox{}^{*}$,
indicating the stationary state, in the
following. 

We  also need the Taylor expansion of the zero-flux condition~(\ref{Knul})~:
\begin{align}
0 = \ & \rho(1-\rho-\sigma) - \sigma\left[1-(\rho+\sigma)^2\right]
- \frac{1}{\LL}\left[\rho \frac{\partial (\rho+\sigma)}{\partial x}
+ \left(1-(\rho+\sigma)^2\right) \frac{\partial \sigma}{\partial x}
\right]
\nonumber \\
&
+ \cdots
\label{taylor_exp_K}
\end{align}

We know from (\ref{s1_ness}) and (\ref{eq_scaling})
that the density of minus particles on the first site $\sigma_1$ is
of order $\mathcal{O}\left(\frac{1}{\LL^2}\right)$.
As in the low density phase there are no steep gradients
(apart from near the right boundary),
density derivatives will also be at most of this order.

We  thus write the densities as
\begin{subequations}
\begin{eqnarray}
\rho &=& \rho^{(1)} + \rho^{(2)} + \rho^{(3)} + \mathcal{O}\left(\frac{1}{\LL^5}\right) \\
\sigma &=& \sigma^{(1)} + \sigma^{(2)} + \sigma^{(3)} + \mathcal{O}\left(\frac{1}{\LL^5}\right)
\end{eqnarray}
\end{subequations}
where we assume that $\rho^{(k)} = \mathcal{O}\left(\frac{1}{\LL^{k+1}}\right)$ and equivalently for $\sigma^{(k)}$.
We will then solve equations 
(\ref{taylor_exp_rho},
\ref{taylor_exp_sig} 
\ref{taylor_exp_K}) order by order (see appendix)
resulting in
\begin{subequations}
\begin{eqnarray}
\rho & = & \frac{\alpha}{p} - \frac{2 r \LL}{p} \left(\frac{\alpha}{p}\right)^2 x
+ \mathcal{O}\left(\frac{1}{\LL^5}\right)
\label{rho_lowdens} \\
\sigma & = & \frac{\alpha}{p}\left[1 - 2 \frac{\alpha}{p}\right] - \frac{2 r \LL}{p} \left(\frac{\alpha}{p}\right)^2 x
+ \mathcal{O}\left(\frac{1}{\LL^5}\right).
\label{sigma_lowdens} 
\end{eqnarray}
\end{subequations}

Figure~\ref{free_flow_dens} shows a comparison between direct simulation
of the MF equations and the prediction (\ref{rho_lowdens}, \ref{sigma_lowdens}),
with excellent agreement. {The lowest order of this calculation recovers
that of \cite{lin_s_a2011}. 
However, when  considering higher order terms in $1/L$ of the densities,
the non-linear terms in  (\ref{taylor_exp_rho},
\ref{taylor_exp_sig}) play a role.}

\section{Analysis of the pulsing phase: the filling stage}
 We now turn to the  analysis of  
the behaviour in the pulsing phase which involves filling and emptying
stages as part of its cycle.
First we show some numerical results and then compare those with analytical expressions.
For the analytical expressions we assume that 
\begin{equation}
p = \mathcal{O}\left(\LL\right), \;\;\;
r = \mathcal{O}\left(\frac{1}{\LL}\right),  \;\;\;\mbox{ and }
\;\;\alpha = \mathcal{O}\left(1\right).
\label{eq_scaling_puls}
\end{equation}
This scaling ensures that we are in the pulsing  phase (see Figure \ref{phase_diagram}).

As can be seen from Figure~\ref{density_2l} (c)  the filling stage 
begins with an empty system  ($\rho$, $\sigma \simeq 0$). Then, as the system rapidly fills, it is divided into
two regions: a low density region to the left with $\rho$, $\sigma
\simeq 0$ and a high density region to the right with $\rho$, $\sigma \simeq 1/2$. 
\begin{figure}[htb]
\begin{minipage}{0.5\textwidth}
\includegraphics[width=\textwidth]{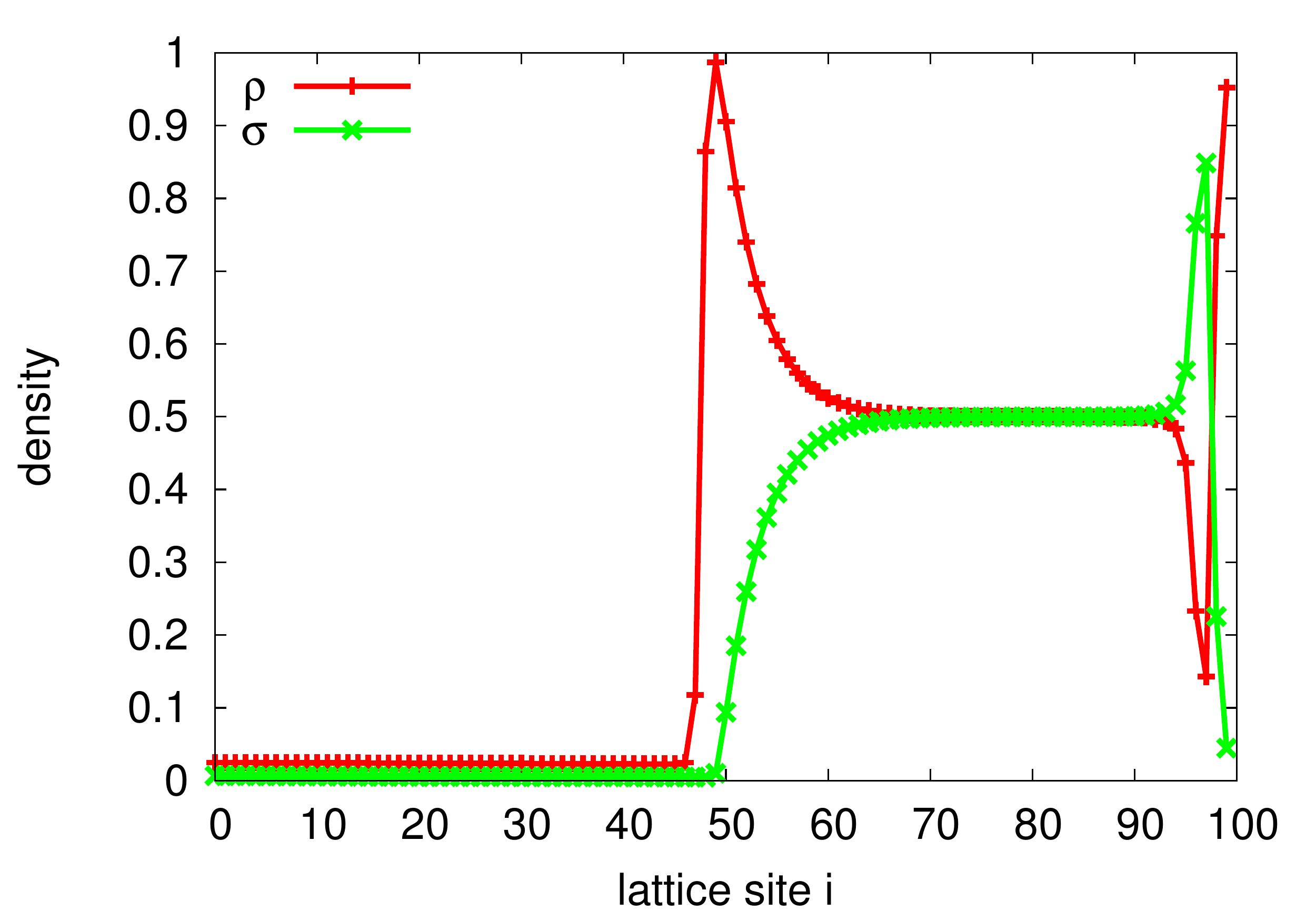} 
\centering(a)
\end{minipage}
\begin{minipage}{0.5\textwidth}
\includegraphics[width=\textwidth]{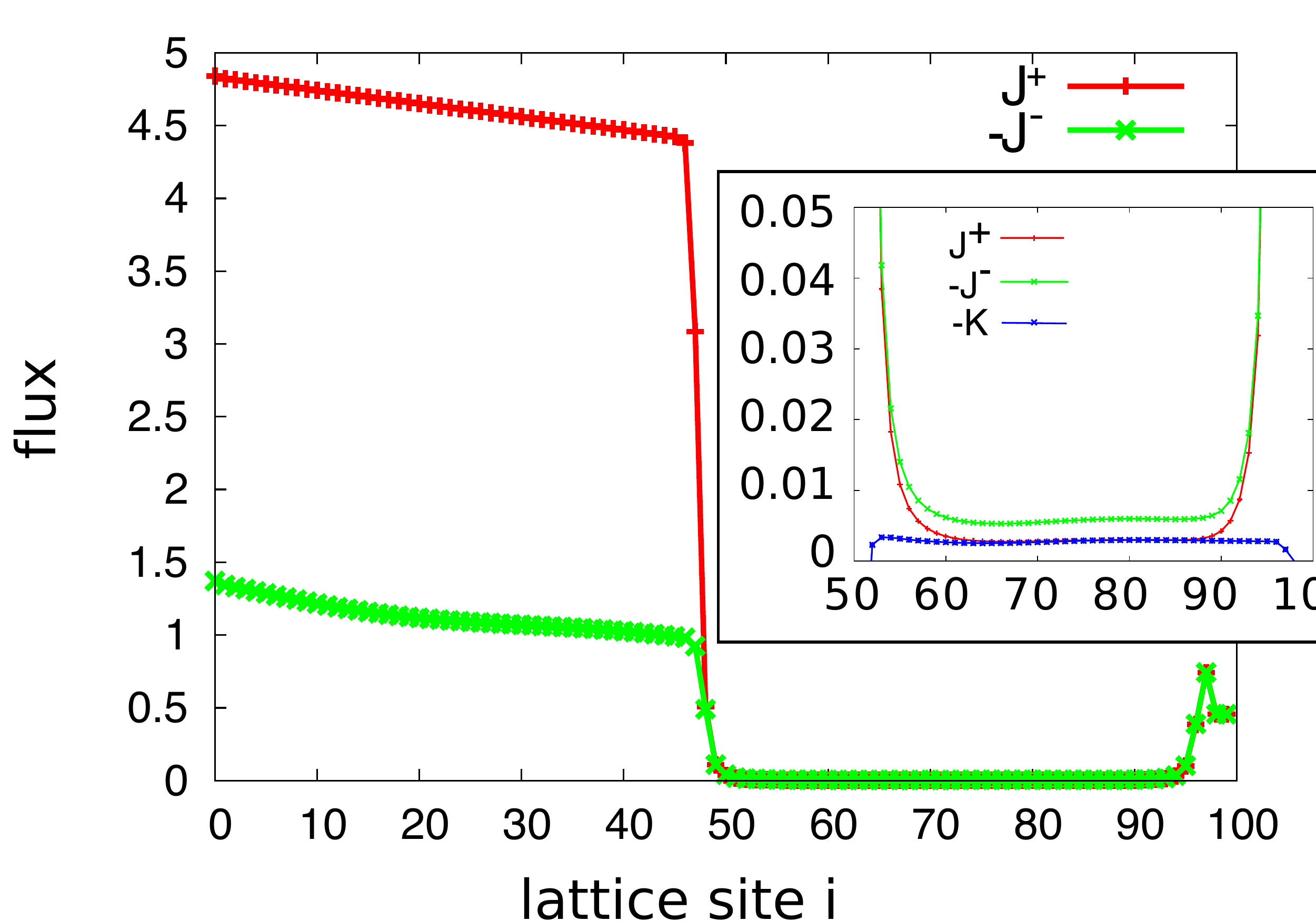}
\centering(b)
\end{minipage}
\caption{Density profiles (a) and fluxes (b) for $\alpha=5$ and $r=0.5$ in the filling stage.
Note that actually we plot $-J^-$ to allow a direct comparison with $J^+$. The 
inset in (b) shows that the fluxes converge towards $J_- = 2J_+$ in the high density
region in the filling stage. 
}\label{filling_profiles}
\end{figure}

In Figure \ref{filling_profiles} (a) the densities $\rho$ and $\sigma$ are shown at
a given time $t$ during the filling stage of the system.
The two regions are separated by a front structure which moves to the
left and
occupies a few lattice sites over which $\rho$ peaks to one and
$\sigma$ dips to zero. 
The peak in $\rho$  corresponds to  plus particles that
cannot enter the high density region to the right.

One also observes a peak in $\rho$ near the wall, as particles
cannot go further to the right.
The structure of this boundary layer is actually more complex 
because $\rho$ and $\sigma$ oscillate out of phase
near the wall while converging 
to a value of approximately one half in the bulk.
Indeed, in this high density region, fluxes are almost
zero. Note, however, that in absolute value, $J_-$ is double  $J_+$,
a feature that will be explained in the next section.

Looking  at the profile of
the flux per bond (Figure \ref{filling_profiles} (b))
in the low density region, it becomes obvious that the positive
flux is (in absolute value) {greater} than the flux out of the system and so the overall density
increases with time until the density is almost one at every site (cf. Figure \ref{density_2l} (b)).

\begin{figure}[htb]
\centering\includegraphics[width = 0.6\textwidth]{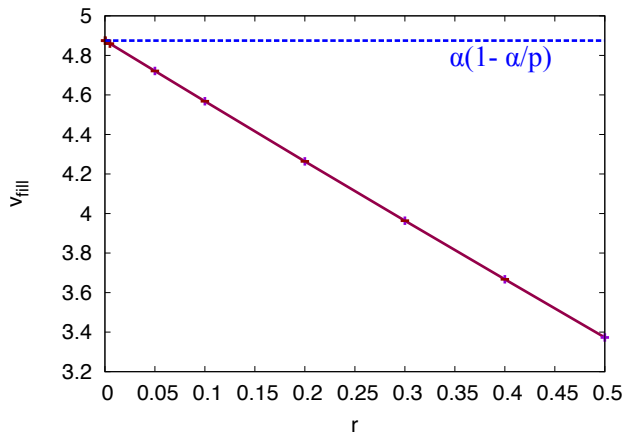}
\caption{$r$ dependence of the filling velocity $\vfi$ when the system is in a pulsing
state (here for $\LL=100$, $p=200$, $\alpha=5$). The dashed blue line shows the
analytic prediction of eq. \eqref{vfill}, valid for small $r$.
}\label{filling_duration}
\end{figure}

As a first step in understanding the filling stage, we wish to
predict the velocity of the moving {shock} front.
To do so, we notice that, in the low density region, the density of minus particles
is much smaller than the density of plus particles. For this reason we
neglect the outflux of minus particles to predict the filling velocity.
If we assume that the density profiles in the low density region
are stationary, so that $K$ is constant in this region,
then the inflow of particles $K_0$
feeds the high density region of density almost $1$, which grows
with velocity $\vfi$. Thus we obtain
\begin{align}\label{vfill}
\vfi= K_0 = \alpha(1-\rho(0)) = \alpha \left(1-\frac{\alpha}{p}\right).
\end{align}
For the scaling given in \eqref{eq_scaling_puls} we have good agreement between the predicted
filling velocity and the numerical results (Figure \ref{filling_duration}). If $r$ increases one observes
deviations from this value. A measurable outflux of minus particles (Figure \ref{filling_profiles} (b))
reduces the filling velocity.  However,
a numerical study (not presented here) of the $\sigma$ density profile in the low density region of the filling phase
reveals that it is not stationary in time.
{Some further effort would be required to take into account this non-stationarity.}

\section{Analysis of the pulsing phase: the emptying stage}

\label{numobs}
As can be seen from Figure~\ref{density_2l} (c) the emptying stage
begins with a full lattice ($\rho$, $\sigma
\simeq 1/2$). Then, as the lattice empties the system is divided into
two regions: a high density region to the left with $\rho$, $\sigma
\simeq 1/2$ and a low density region to the right with $\rho$, $\sigma
\ll 1$. The two regions are separated by a front structure which
occupies a few lattice sites over which $\sigma$ peaks to one. The
front structure moves to the left with a speed very much less than the
speed of the front in the filling stage (see previous section).

\begin{figure}[h!]
\begin{minipage}{0.5\textwidth}
\centering (a)
\includegraphics[width=\textwidth]{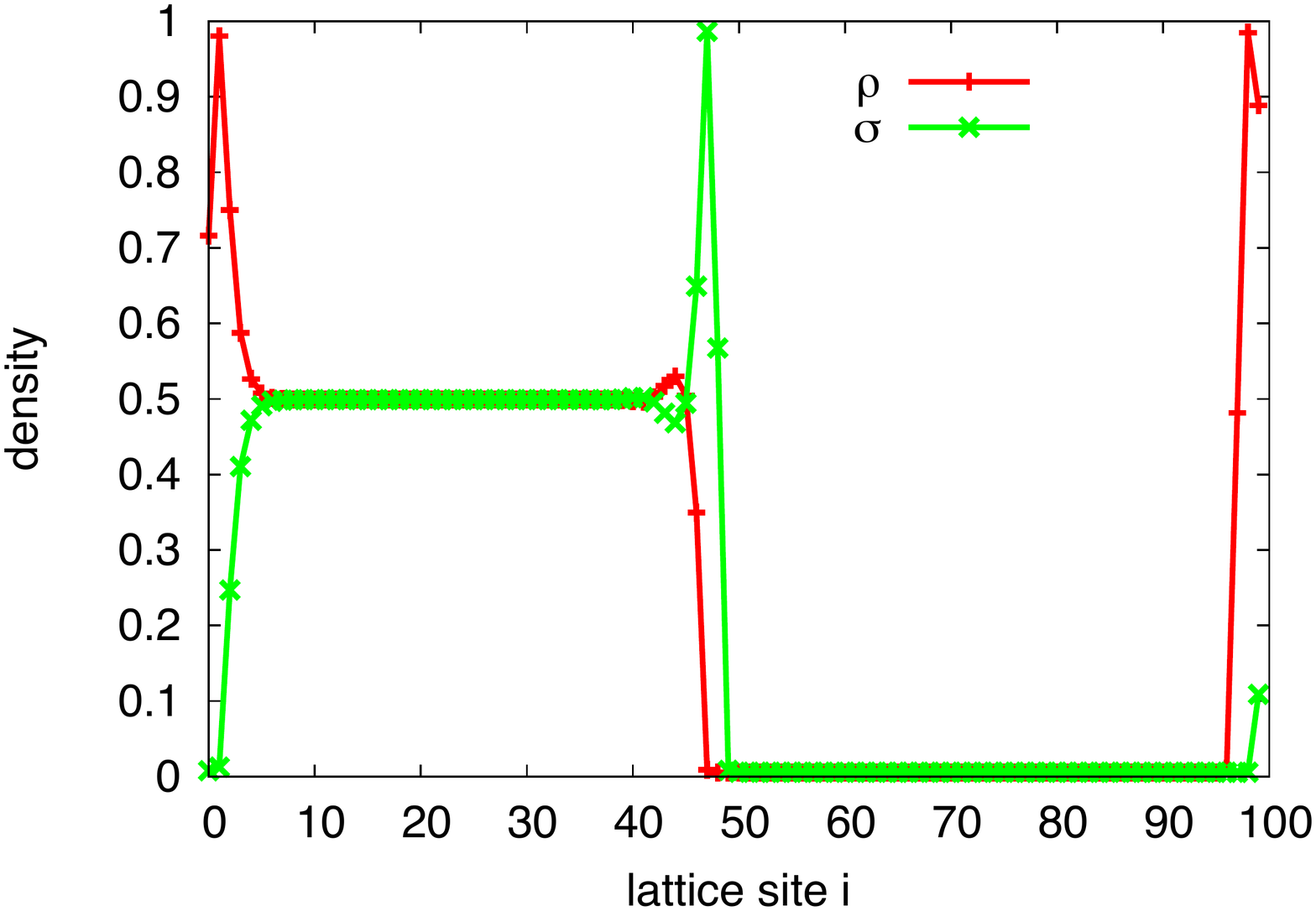}
\end{minipage}
\begin{minipage}{0.5\textwidth}
\centering (b)
\includegraphics[width=\textwidth]{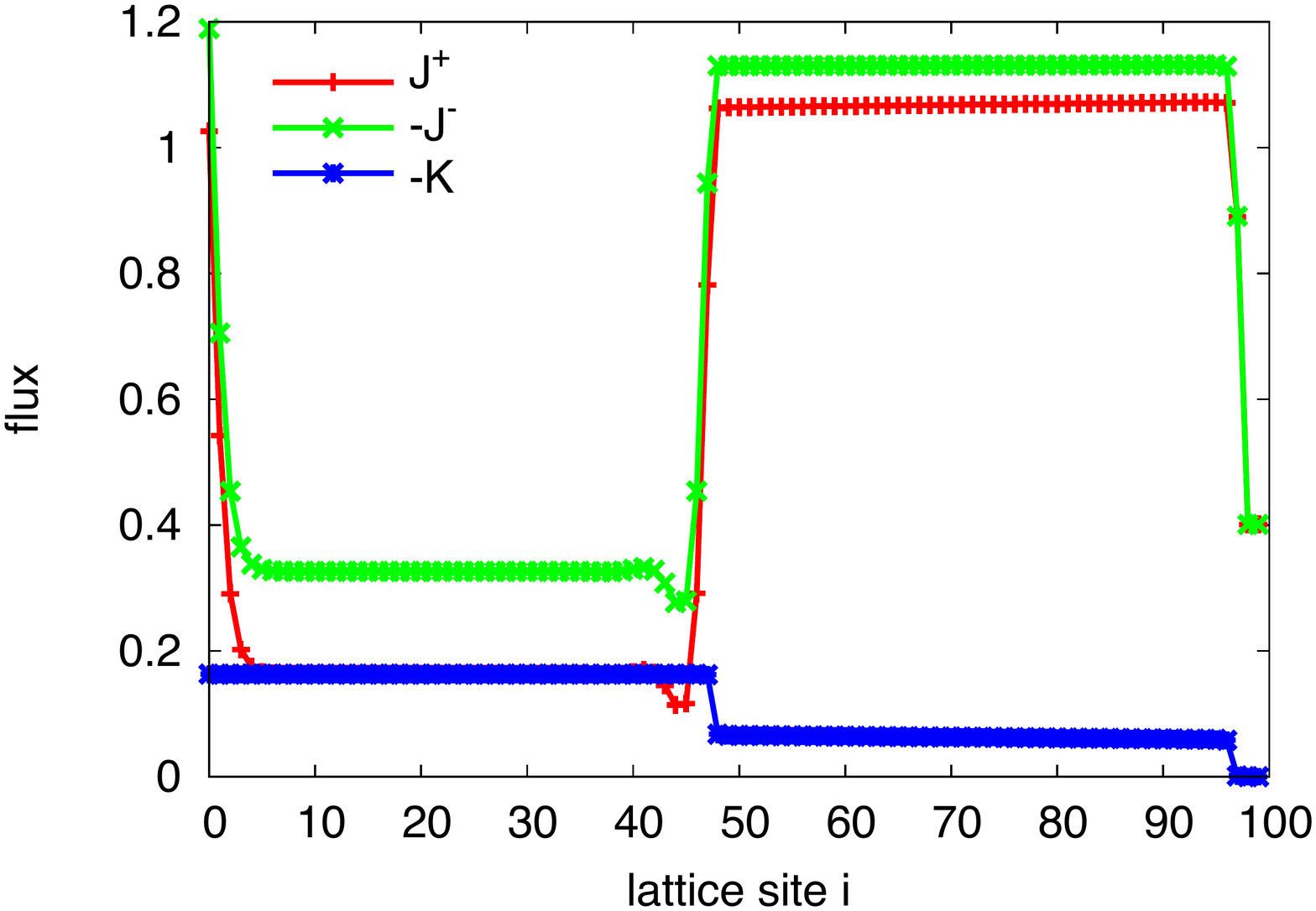}
\end{minipage}
\caption{Density profiles (a) and fluxes (b) in the emptying stage for $p=200$, $\alpha=5$ and $r=0.5$. 
Note that actually we plot $-J^-$ and $-K$ to allow a direct comparison with $J^+$.
{The densities $\rho$ and $\sigma$ for this parameter set are
typically $0.4992...$ in the plateau of the dense region,
and $6\cdot{10^{-3}}$ in the low density phase
(see Figs~\ref{logplot} and~\ref{oscizoom}).
}
}
\label{alpha5r05}
\end{figure}

In Figure \ref{alpha5r05} we examine in more detail the density profiles
and particle fluxes in the two regions. Figure \ref{alpha5r05} (a) shows the densities $\rho$ and
$\sigma$ at a {single moment in time}. 
From the values {of the densities} given in the caption, we already see that the deviations
from $\rho = \sigma = \frac{1}{2}$ in the dense region on the left-hand side of the shock and from $\rho = \sigma = 0$ in the dilute
region on the right-hand side are very small.

In Figure \ref{alpha5r05} (b) $|J^-|, \ J^+$ as well as $|K|$, the
modulus of the net mass current, are shown. 
From the numerical data we find two interesting
facts: First, in the dense region the flux of minus particles is
double the flux of plus particles (in absolute value). We shall
explain this fact in the next section. Furthermore,
$K$ is constant in the dense region whereas it 
decreases slightly (in
absolute value) with increasing $i$ in the dilute region.

The first observation implies that $K$ is negative, confirming that
there is a net mass flow out of the system at the left end. The net
mass flow is still negative but small in the low-density right-hand
region although the net individual currents are larger than in the
dense region. The linear increase in $K$ in the low density region
corresponds to a linear decrease in $\rho$ in this low density region.

\subsection{Shock dynamics}
\label{sec:shock}

We idealise the moving shock seen in Figure~\ref{alpha5r05} (a)
as a discontinuity (possibly with a complex shape, as long as it is of
finite spatial extent)
separating two  domains of constant densities.
Our aim is to predict the shock velocity through Rankine-Hugoniot equations,
which give a relation between the densities and the fluxes
on both sides of the shock.
In doing so, we  implicitly assume that mass is conserved
for each species, so that the
source terms $\pm r(\rho_i - \sigma_i)$
corresponding to transmutations
will not be taken into account.

We denote the fluxes of plus and minus particles on both
sides of the shock by
\begin{subequations} \label{fluxdef2}
\begin{align}
\label{fluxdef2:pl}
&J^+_X = p \rho_X(1-\rho_X-\sigma_X) \\
\label{fluxdef2:mi}
&J^-_X = -p\sigma_X[1-(\rho_X+\sigma_X) ^2]\;,
\end{align}
\end{subequations}
where the index $X= \{ \Le,\Ri\}$ 
refers to the
left{-} and right{-}hand side of the shock, respectively.
Furthermore, we define the net current $K$ on each side of the shock as
\begin{align}
K_X & \equiv  J^+_X + J^-_X.
\end{align} 

In a uniform domain (flat density profiles),
the equations (\ref{taylor_exp_rho},\ref{taylor_exp_sig})
become
\begin{equation}
\frac{\partial \rho}{\partial t} =
- \frac{\partial \sigma}{\partial t} =
- r \left( \rho - \sigma\right)
\label{source}
\end{equation}
Thanks to the source term on the right\sarahcol{-}hand side of (\ref{source}), the system relaxes
towards a state in which $\rho=\sigma$,
which is indeed what is observed at lowest order in the numerics.
We shall thus assume that $\rho=\sigma$ in each domain
on each side of the shock.
We know from the numerics that these values are close 
to $1/2$ and $0$, respectively.
More precisely,
using the knowledge gained from the numerical results for the densities (Figure \ref{alpha5r05}(a)) we  make the following assumption
\begin{align} \label{approx}
\rho_\Le = \sigma_\Le = \frac{1 -\eta_\Le}{2}
\end{align}
with $\eta_\Le \ll 1$ (in Figure~\ref{alpha5r05}(a),
we find $\eta_\Le = 1.63\cdot10^{-3}$).

Inserting this in the flux definitions \eqref{fluxdef2} we find
\begin{subequations}
\label{approx_J}
\begin{align}
J_\Le^+ = \frac{p\eta_\Le}{2} + \mathcal{O}(\eta_\Le^2) \\
J_\Le^- = -p{\eta_\Le} + \mathcal{O}(\eta_\Le^2)
\end{align}
\end{subequations}
which explains 
why the  flux of minus particles is double that
of plus particles to lowest order in $\eta_\Le$. 


The Rankine-Hugoniot relations are given by
\begin{subequations}
\begin{align}
J_\Le^+ - J_\Ri^+ = v(\rho_\Le - \rho_\Ri) \\
J_\Le^- - J_\Ri^- = v(\sigma_\Le - \sigma_\Ri)
\end{align}
\end{subequations}
where we assume that the front for both species moves at the same speed $v$.
These equations simply express mass conservation for each species at
the shock.
Using the expressions (\ref{approx}-\ref{approx_J}) and observing, furthermore,
that $\sigma_\Ri$ and $\rho_\Ri$ are small compared to $\rho_\Le$, $\sigma_\Le$ so that we can neglect second order terms in $J_\Ri^\pm$, we obtain
\begin{subequations}\label{RH2}
\begin{align}\label{RH2:1}
p\left(\frac{1}{2}\eta_\Le - \rho_\Ri\right) &=  v\left(\frac{1}{2} - \frac{\eta_\Le}{2} -\rho_\Ri\right) \\
\label{RH2:2}
p\left(-\eta_\Le + \sigma_\Ri\right) &= v\left(\frac{1}{2} - \frac{\eta_\Le}{2} - \sigma_\Ri\right).
\end{align}
\end{subequations}
We can now eliminate $v$ from the equations \eqref{RH2}
by assuming that $\sigma_\Ri =\rho_\Ri$ so as to  obtain,  to
leading order, the relation
\begin{align}\label{RH3}
\rho_\Ri = \sigma_\Ri = \frac{3}{4} \eta_\Le
\end{align}
\begin{figure}[tb]
\begin{minipage}{0.5\textwidth}
\centering
\includegraphics[width = \textwidth]{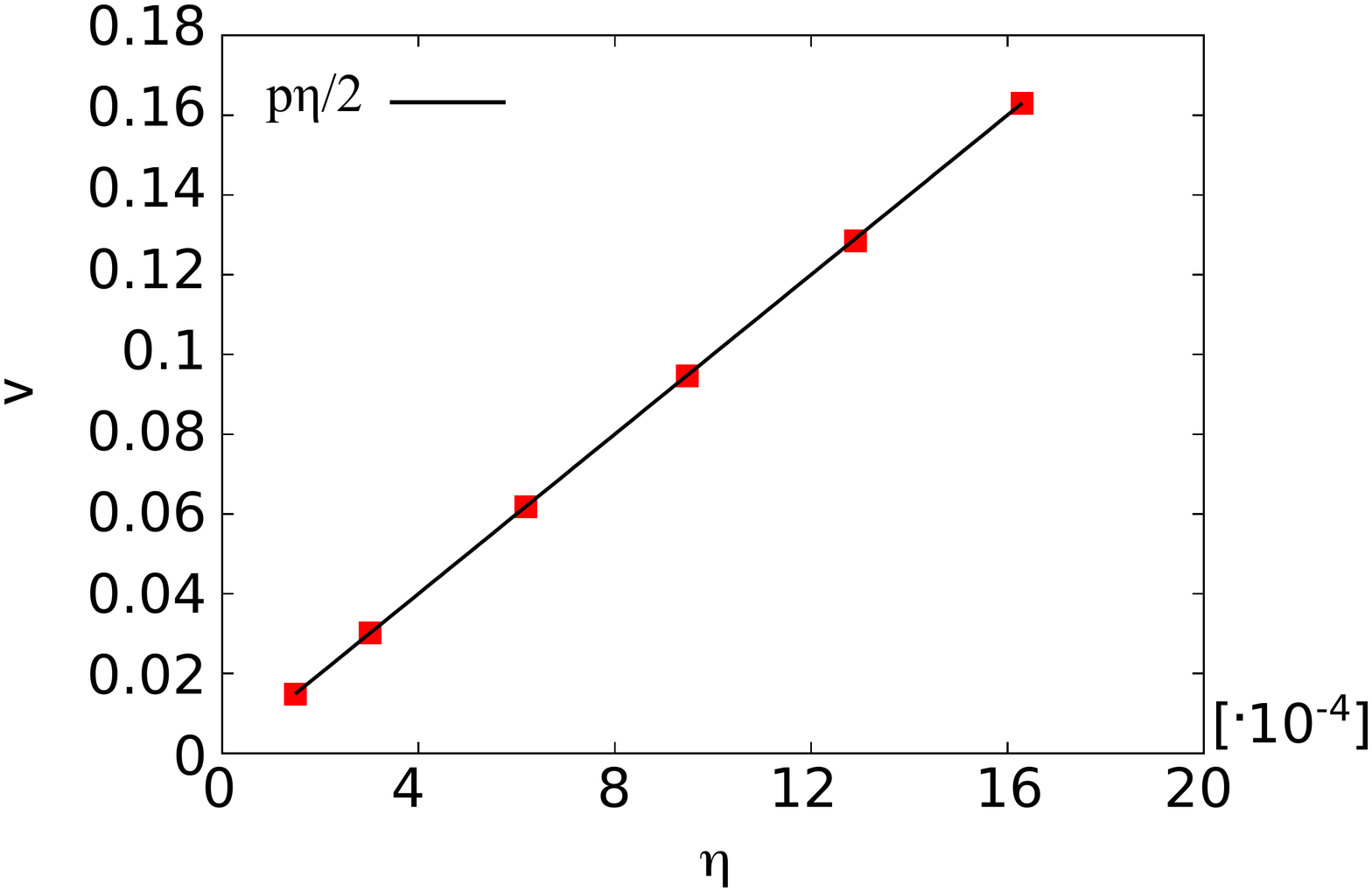}\\
 (a)
\end{minipage}
\begin{minipage}{0.5\textwidth}
\centering
\includegraphics[width = \textwidth]{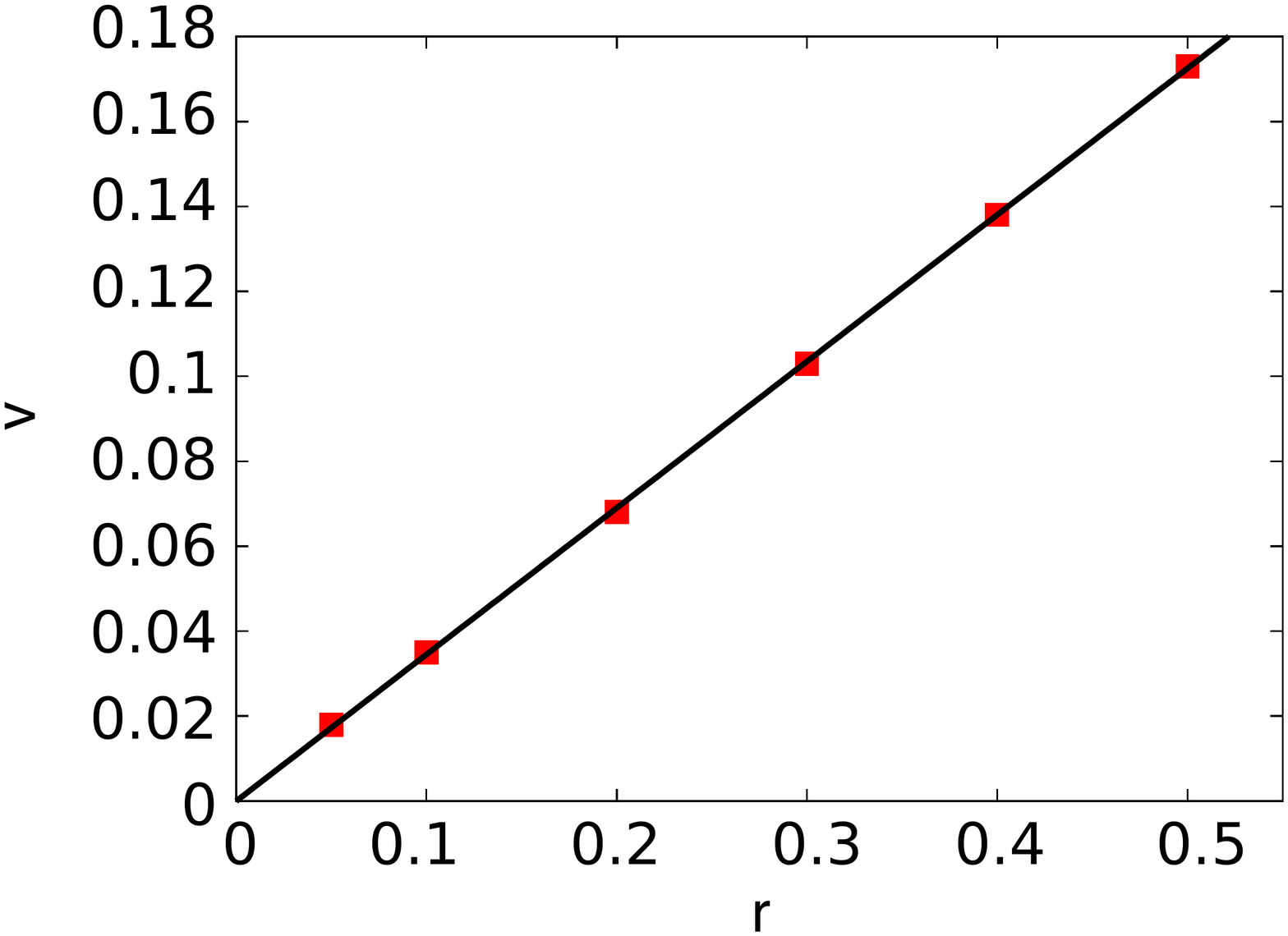}\\
 (b)
\end{minipage}
\caption{Dependencies of the emptying front velocity $v$. (a) $v$ against
the deviation $\eta_\Le$ from density 1/2 
in the dense region and (b) $v$ against the 
transmutation rate $r$. 
The points on both figures correspond to the same simulations so that the
value of $\eta_\Le$ for the parameter values used throughout this paper ($r=0.5$)
is  $\eta_\Le=1.63\cdot 10^{-3}$.
In
(a) the line corresponds to the Rankine-Hugoniot expression given in
eq.~\eqref{RHC} and shows very good agreement.
Furthermore, we find that the velocity is also proportional to $r$ (the line in (b) is a fit with slope $0.345$).\\
Parameters: $\alpha=5$, $p=200$.

}\label{velof}
\end{figure}
%
and an expression for the front velocity
\begin{align}\label{RHC}
v = -\frac{p}{2}\eta_\Le.
\end{align}
In Figure \ref{velof} we test the predicted relation (\ref{RHC})
numerically and find 
that  indeed  $\eta$ and $v$ are
proportional to each other, with a proportionality coefficient
well predicted by the Rankine-Hugoniot prediction (\ref{RHC}).
Furthermore, we checked that $\eta_\Le$ does not depend on the system size
(see for example Figure~\ref{logplot}~(c)).
However, we do not have  a prediction for the value of $\eta_\Le$.
In  Figure~\ref{velof}~(b), we observe that $v$---and thus
also $\eta_\Le$---is proportional to $r$.


In the Rankine-Hugoniot calculation, we have assumed constant
domains on both sides of the shock.
We shall see now that this is not fully {accurate}, and that
a more careful {study} reveals some oscillations,
both in the dense and in the dilute region.

\subsection{Dense region}

In the dense region, to leading order, the density of both species are
equal
and  far away from the front are approximately equal to one half.

\subsubsection{Oscillations in the dense region}

\begin{figure}[htb]
\includegraphics[width=\textwidth]{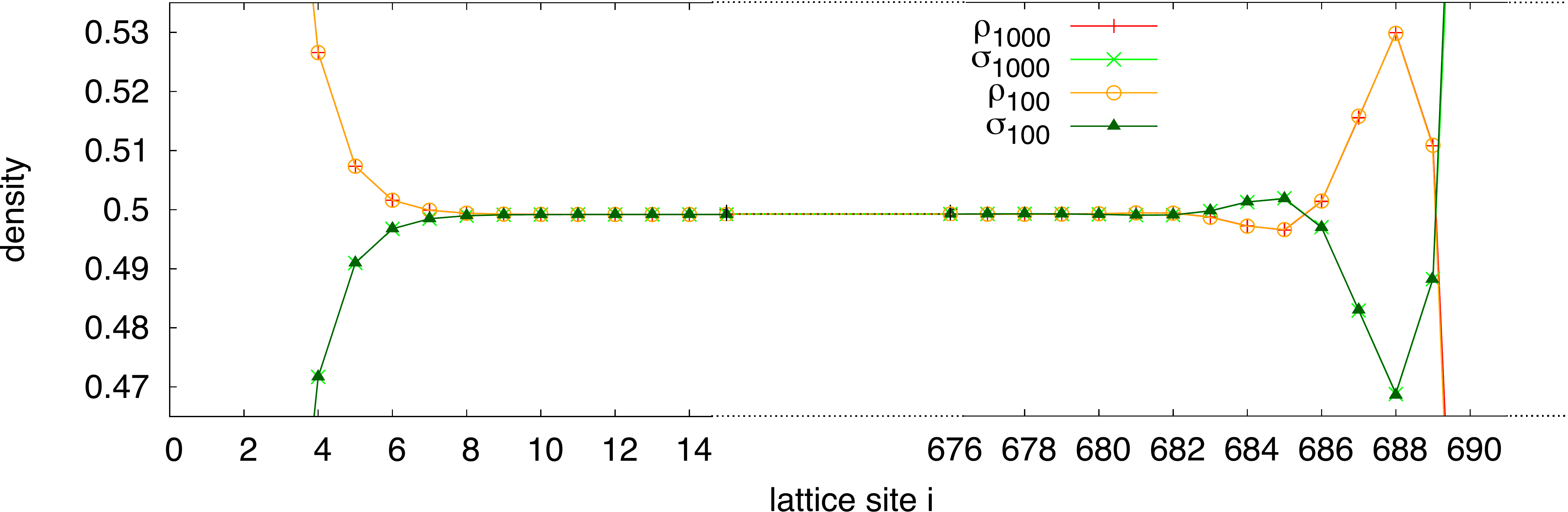}\\
\centering 
(a)\\
\begin{minipage}{0.49\textwidth}
\includegraphics[width=\textwidth]{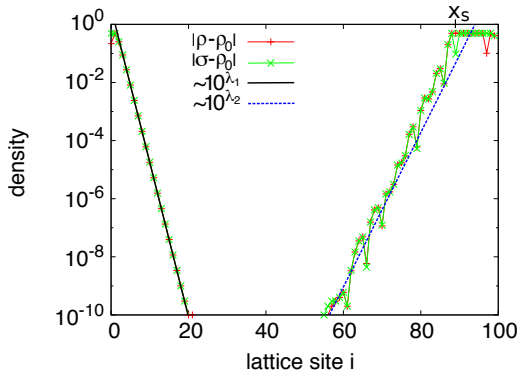}\\
\centering
(b)
\end{minipage}
\begin{minipage}{0.49\textwidth}
\includegraphics[width=1.05\textwidth]{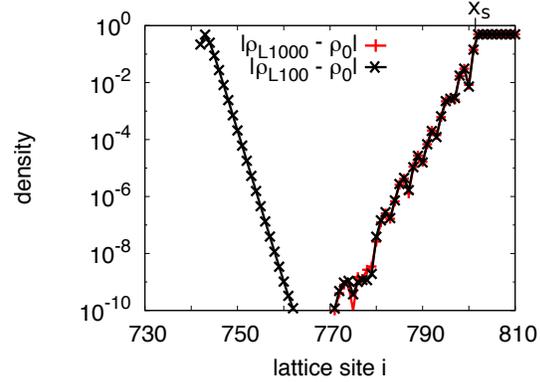}\\
\centering
(c)
\end{minipage}
\caption{Density profiles $\rho$ and $\sigma$
in the dense region of the emptying stage. 
(a) Zoom around density $1/2$ on the density profiles $\rho$ and $\sigma$,
for $\LL=100$ and $\LL=1000$. Only the sites near the boundaries
of the dense region are shown (with a translation of the right sites
in the case $L=100$, in order to overlap with the $L=1000$ data).
One sees the decay of density perturbations {to the bulk values $(\rho_0, \sigma_0)$} at the entrance
and the oscillatory behaviour on the {right of the figure} (left-hand side of 
the shock position $x_s$). 
(b) $\log(|\rho - \rho_0|)$ and $\log(|\sigma - \sigma_0|)$ for $\LL=100$, the straight lines show the results 
obtained by the stability analysis of section~\ref{sec_stability}. 
(c) $\log(|\rho-\rho_0|)$ for $\LL=100$ (black) and $\LL=1000$ (red)
(the $\LL=100$ data points are translated horizontally in order
to have the same $x_s$ position as for $\LL=1000$).
Parameters: $\alpha = 5$, $r = 0.5$, $p=200$. For this parameter set,
we use the same $\rho_0 = \sigma_0 =0.499182602$ for the two different lattice lengths.
} \label{logplot}
\end{figure}

Inspecting Figure \ref{logplot}(a),
we observe that in the bulk of the dense region,
both densities reach a plateau value
$\rho_0 = \sigma_0 = (1 - \eta_\Le)/2$.
On both sides of the dense region, we observe
deviations from these plateau values.
The semi-logarithmic plots
$\log(|\rho - \rho_0|)$ and $\log(|\sigma - \sigma_0|)$
in Figure \ref{logplot}(b) show that at the entrance,
the deviations from the plateau densities decay exponentially
when the particles penetrate the system.
On the right\sarahcol{-}hand side of the dense region,
we find some spatial oscillations in the density profiles
that are also damped  exponentially in the bulk.

In the simulations (see Figure \ref{logplot}(a)),
the oscillations in the dense region to the left-hand side
of the shock have a wavelength of about seven
sites both for $\LL=100$ and $\LL=1000$.
Furthermore, we observe that $\rho$ and $\sigma$ oscillate
in opposite phase, with an almost perfect symmetry. Indeed,
$\rho_0 - \rho$ and $\sigma-\rho_0$ overlap up to at least $10^{-6}$.
Again, this holds both for $\LL=100$ and $\LL=1000$. 

In Figure \ref{logplot}(c)
we superimpose the profiles 
$\log(|\rho-\rho_0|)$ 
for two different lattice
lengths $\LL=100$ and $\LL=1000$. 
We can see that the oscillations on the left-hand side of
the shock appear to have the same exponential decay for the two lattice
lengths, the same wavelength and even the same structure.
A zoom in Fig \ref{logplot}(a) (not shown) shows also that
there is no significant
difference between the two systems sizes either for the wavelength  or
for the amplitude of the oscillations.
Thus, the oscillatory behaviour appears to be
independent of system size.
The exponential decay at the entrance of the dense region
is also independent of system size.

Measurements of the slopes of $\log(|\rho-\rho_0|)$ for $\LL=100$ and $\LL=1000$ in Figure \ref{logplot}(c) give at the entrance a 
decay constant $\simeq -1.07$ and
close to the shock a decay constant $\simeq 0.8$. 

We shall see in section~\ref{sec_stability}
how a stability analysis can explain these observations.

\begin{figure}[htb]
\centering
\includegraphics[scale=0.4]{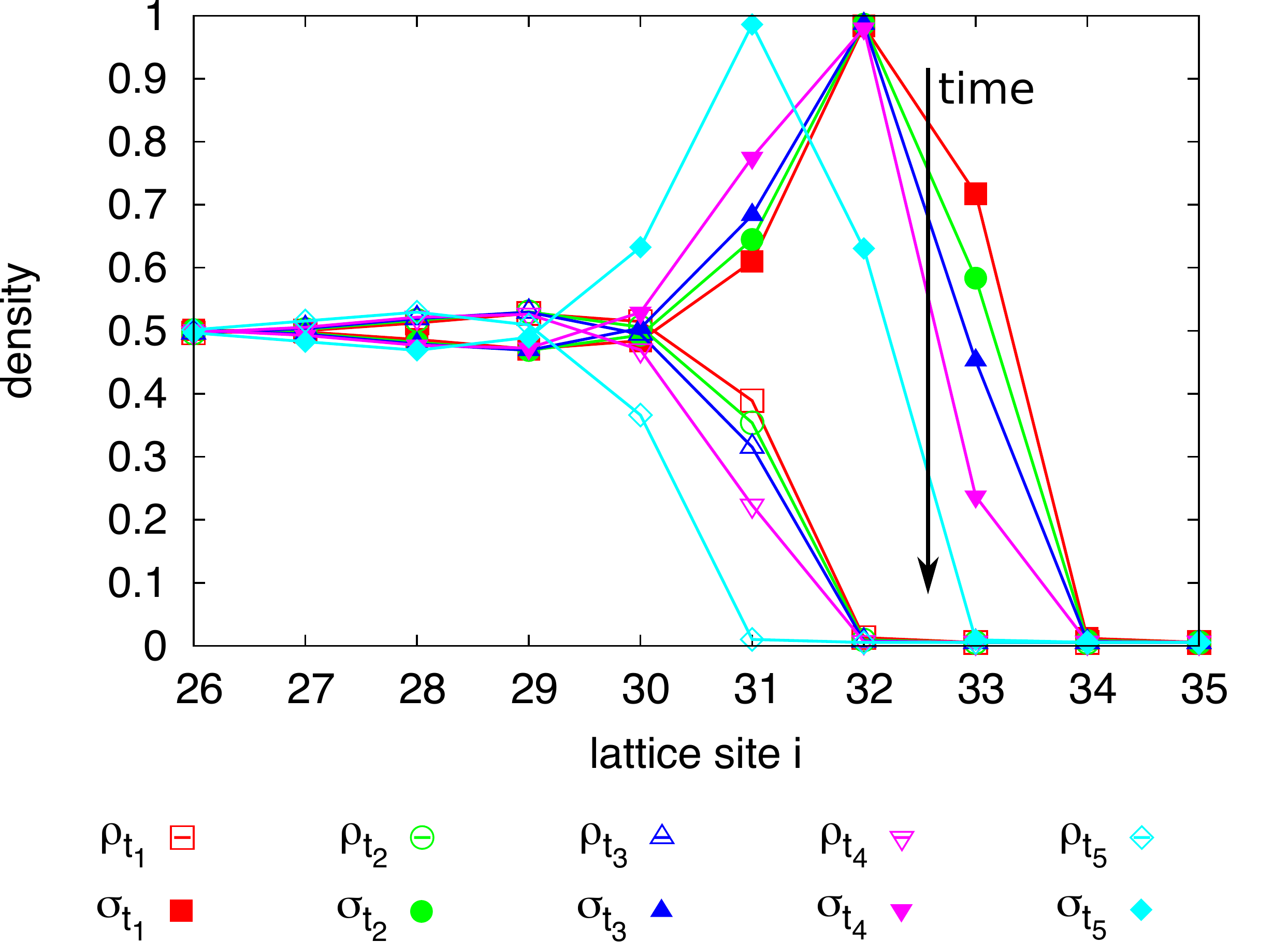}
\caption{Illustration of the shock moving by one site. Given are
the density profiles for $\rho$ (empty symbols) and $\sigma$ (filled symbols) at different times. 
We can observe how the density profiles are
deformed when the shock moves one step to the left.
The profile at $t_5$ is the same as at $t_1$ translated by one lattice unit
to the left. Here, $\alpha=5$, $r=0.5$, $L=100$,
and we find $t_5-t_1= 6.05$ time units.
The time interval between two successive profiles
is $(t_5-t_1)/4$.
}
\label{profile_move}
\end{figure}

\clearpage
\subsection{Dilute region}\label{sec:dilute}
In the emptying stage of the pulsing phase there is a dilute region to
the right of the moving shock (see Figures \ref{alpha5r05}). The analysis
of Section \ref{sec:shock} assumes equal low densities
$\rho_\Ri=\sigma_\Ri$ to the right of the shock.
However, in this section we carry out a more detailed numerical study that
reveals that, though they are on average of the same order,
the densities $\rho_\Ri$ and $\sigma_\Ri$ actually differ
and oscillate in time.

\begin{figure}[tb]
\centering
\includegraphics[angle=270,width=0.7\textwidth]{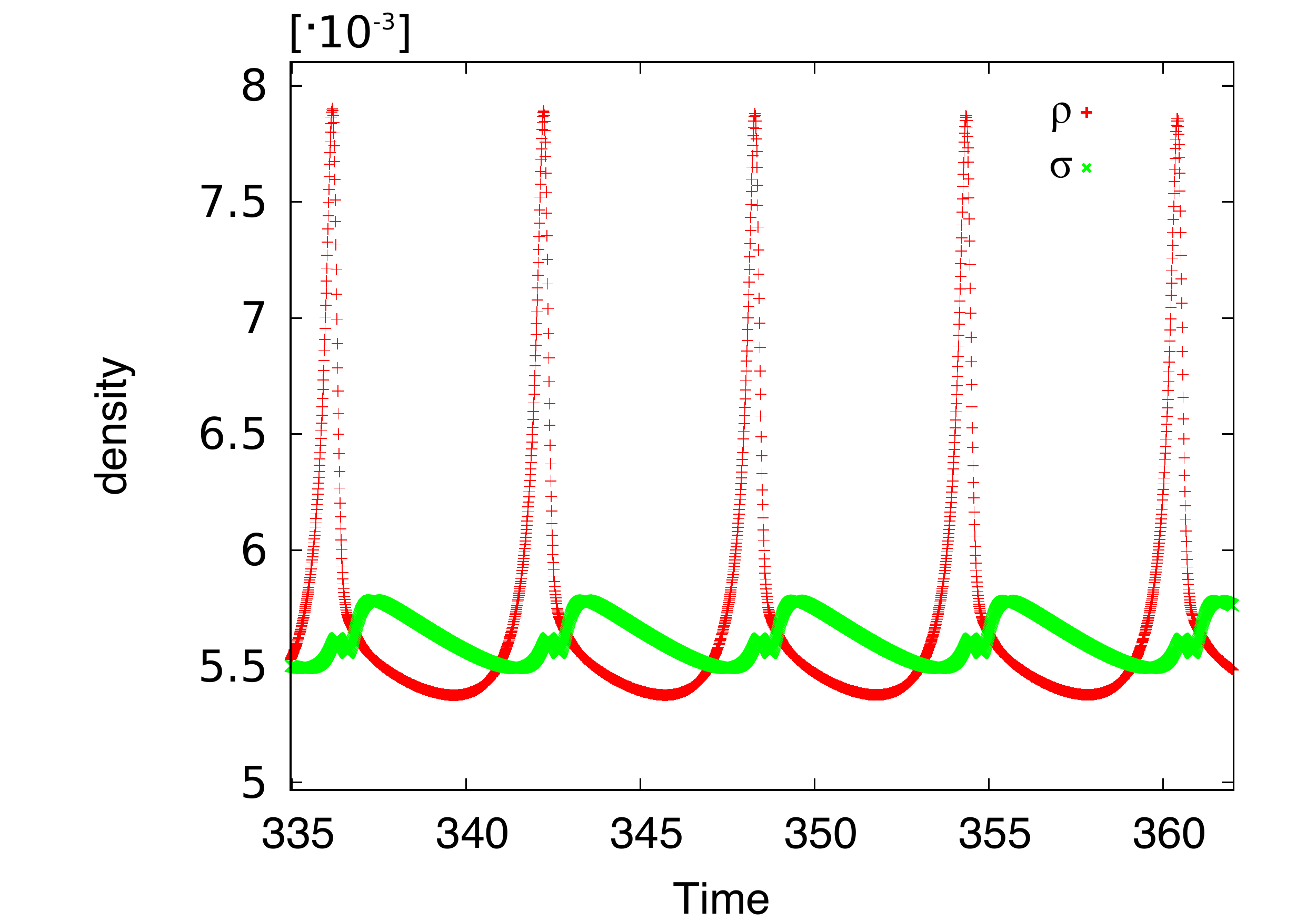}
\caption{Temporal oscillations of $\rho_\Ri$ (red)
and $\sigma_\Ri$ (green) at a fixed lattice site $i$ in the
dilute region of the emptying state.
Both densities
exhibit non-sinusoidal oscillations.\\
Parameters: $p=200$, $\alpha = 5$, $r=0.5$.
}
\label{oscizoom}
\end{figure}

{Figure~\ref{oscizoom} illustrates the temporal variations
of both densities at one given lattice site~$i$ located 
in the dilute region.
We observe fast oscillations in $\rho_\Ri$ and $\sigma_\Ri$
as a function of time.
Neither $\rho_\Ri$ nor $\sigma_\Ri$
change in a sinusoidal way, rather they  show highly non-linear behaviour.
}

{
For the parameter set used in Figure~\ref{oscizoom},
the typical value around which the densities oscillate is $5.5 \times {10}^{-3}$,
a value which differs from
the value predicted by the Rankine-Hugoniot relation (\ref{RH3}),
$\rho_\Ri = \sigma_\Ri = \frac{3}{4} \eta_\Le = 1.2 \times {10}^{-3}$.
This is not surprising, as
obviously, the assumption that $\rho_\Ri$ and $\sigma_\Ri$ are
equal and uniform is too naive.
Consideration of the source terms
(which have been ignored in the analysis of Section \ref{sec:shock})
and the complex structure of the density profiles would be crucial
to find a solution in this region.
It seems out of reach to derive {the profiles} analytically,
but several interesting observations can be made.
}

\begin{figure}[htb]
\begin{minipage}{0.5\textwidth}
 \includegraphics[width=\textwidth]{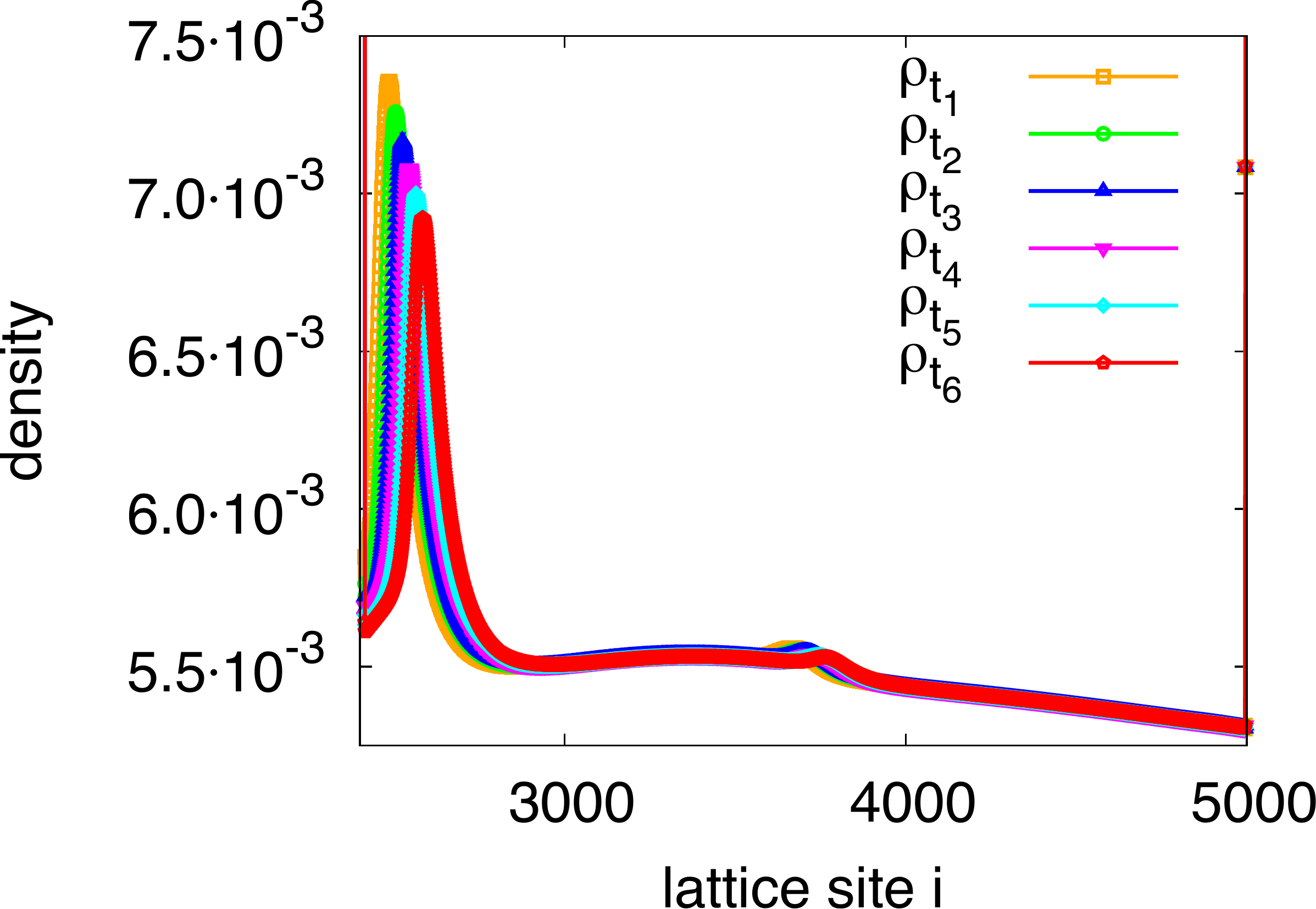}
\end{minipage}
\begin{minipage}{0.5\textwidth}
 \includegraphics[width=\textwidth]{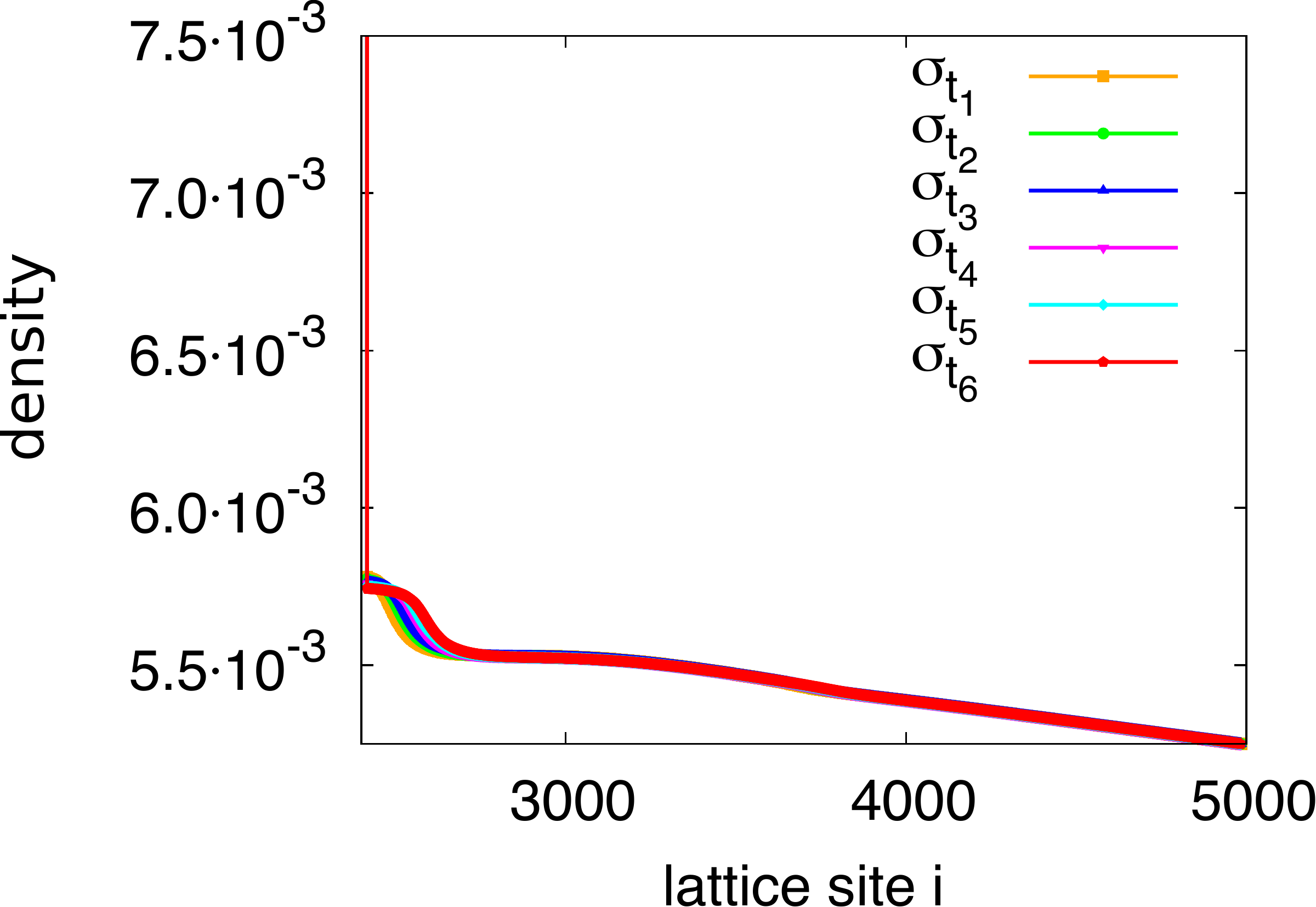}
\end{minipage}
\caption{The two densities in the dilute region of the emptying stage, (a)
$\rho$ and (b) $\sigma$ at various equidistant times.
The time interval shown $t_6-t_1 = 0.6$ time units is less than a period
in Figure~\ref{oscizoom}, or equivalently, less than the time interval shown
in Figure~\ref{profile_move}.
One observes, close to the shock, a {microscopic} pulse propagating to the right
with a decaying amplitude as it propagates.
One can also guess the presence of another {microscopic} pulse around $i=3700$
previously emitted and already strongly damped.
The parameters values are  $\alpha = 5$, $r=0.5$, $p=200$.
}\label{zoom}
\end{figure}

{
The period of the oscillations in Figure \ref{oscizoom}
is about 6 time units, which
corresponds to the {time period} for the shock to move
from one lattice site to the next one, as shown
on Figure~\ref{profile_move}. Thus, this oscillatory behaviour seems directly related
to the hopping of the shock on the discrete lattice.
We checked that the behaviour was robust to
changes in the time-step of the  numerical iteration scheme used to
solve the MF equations.

In Figure \ref{zoom} we plot the density profiles in the dilute region
at closely separated time intervals.
One sees that the shock  {emits microscopic} density pulses which correspond to the
density peaks at a given site seen in Figure~\ref{oscizoom}.
The steepness of the peaks is to be related to the
rather abrupt way in  which the shock moves from one site to the next
(see Figure~\ref{profile_move}).
}


\section{Stability analysis}
\label{sec_stability}

In this section we consider perturbations about constant density
solutions of the MF equations within a linear stability
analysis.  In many works~\cite{evans2011,curatolo2016} it has been shown that a linear analysis of
MF equations of the form of eq. (\ref{perturb}) can give insight into the
full solutions. 

\subsection{Perturbation about constant profiles}

The discrete bulk equations (\ref{bulk_2l_p1}-\ref{bulk_2l_s1})
accept as a stationary solution any constant profile $(\rhostar,\sigstar)$
provided that
\begin{equation}
\rhostar = \sigstar \le \frac{1}{2}
\end{equation}
where the last inequality comes from the fact that the total
density cannot be larger than one.

We {now wish} to perturb this solution and to explore its linear stability.
We denote the time-dependent perturbation about
a constant profile $(\rhostar,\sigstar)$ with $\rhostar =
\sigstar$
\begin{subequations}\label{perturb}
\begin{align}
\label{perturb_1}
\rho_i(t) & = \rhostar   + A(t) \Lambda^i \\
\label{perturb_2}
\sigma_i(t) & = \rhostar + B(t) \Lambda^i \;.
\end{align}
\end{subequations}


On linearising in $A(t)$ and $B(t)$,  the bulk equations (\ref{bulk_2l_p1},\ref{bulk_2l_s1})  become
\begin{subequations}\label{bulk_perturb}
\begin{align}
\label{bulk_perturb_1}
\Lambda^i\frac{\text{d} A}{\text{d}t}   = & 
p \Lambda^i \left(\Lambda-1\right)
\left[(A+B)\rhostar-A(1-2\rhostar)\frac{1}{\Lambda}\right]
-r(A-B)\Lambda^i
\\
\Lambda^i \frac{\text{d}B}{\text{d}t}  = &
p \Lambda^i \left(\Lambda-1\right)
\left[(1-4{\rhostar}^2)B - 4{\rhostar}^2 (A+B)\frac{1}{\Lambda}\right]
+r(A-B)\Lambda^i
\label{bulk_perturb_2}
\end{align}
\end{subequations}
where it should be noticed that a factor $(\Lambda-1)$ 
could be factorised in the flux terms.
These equations may be written in matrix form as
\begin{equation}
\frac{1}{p} \frac{\text{d} }{\text{d}t} \left(
\begin{array}{c}A\\B \end{array}
\right) =
M
\left(
\begin{array}{c}A\\B \end{array}
\right)
\label{eqM}
\end{equation}
where the two-by-two matrix $M$ is defined by
\begin{equation}
M = 
\left(
\begin{array}{lr}
\left(\Lambda-1\right)
\left[ \rhostar - \frac{1-2\rhostar}{\Lambda}\right]
-\frac{r}{p} \;\; 
&
\left(\Lambda-1\right) \rhostar + \frac{r}{p} \\
\left(\Lambda-1\right)\left[ \frac{-4{\rhostar}^2}{\Lambda} \right]
+ \frac{r}{p}
&
\left(\Lambda-1\right)
\left[ 1-4{\rhostar}^2- \frac{4{\rhostar}^2}{\Lambda}\right]
-\frac{r}{p} \;\; 
\end{array}
\right)
\end{equation}
The time dependence of the amplitudes $A(t)$, $B(t)$
is thus determined by the eigenvalues $\omega(\Lambda,\rhostar)$
of the matrix $M$ which 
are functions of $\Lambda$ and $\rhostar$.

The equation for the eigenvalues of $M$
is quadratic in $\omega$ and is fourth order in $\Lambda$,
and it is  in general difficult to write the solutions in a simple
form.  The solutions for $\omega$ provide
information about the temporal stability while the  $\Lambda$
dependence governs
the behaviour of spatial perturbations. 
In the following we will consider two simple relevant  cases separately, first
by assuming a homogeneous  perturbation in space ($\Lambda=1$) and second by assuming
a stationary but spatially varying perturbation ($\omega = 0$).






\subsection{Temporal stability of plateau solutions}

First we  analyse whether plateau (constant density) solutions
are dynamically stable.
For this reason we choose $\Lambda=1$ which gives us an easy equation for the eigenvalues $\omega$ namely
\begin{align}
\omega\left(\omega  + \frac{2r}{p}\right) = 0
\end{align}
with the obvious roots
\begin{align}
\omega_1 = 0 \ \ \  ; \ \ \  \omega_2 = -\frac{2r}{p}.
\end{align}
The corresponding eigenvectors are given by 
\begin{align}
v_1 = \begin{pmatrix}
1 \\ 1
\end{pmatrix} \ \ \  ;  \ \ \ v_2 = \begin{pmatrix}
1 \\ -1
\end{pmatrix}\;.
\end{align}
The first eigenvector corresponds to a perturbation
consisting {of a  uniform and equal} shift of the two densities.
The fact that $\omega_1=0$ confirms that any constant profile
with $\rho=\sigma$ is a stationary state.

The second eigenvector gives the same {shift} to both densities
but with opposite sign. In this case the perturbation decreases
exponentially with time, and both densities 
converge towards the constant value $\rhostar$.
The system is thus \textit{dynamically} stable
{with respect to} this type of perturbation.

For a combination of these shifts, densities will converge,
but towards $(\rho+\sigma)/2$ and not towards $\rhostar$.
From this
we can conclude that the system is \textit{marginally} stable 
with respect to
uniform shifts in the densities.

\subsection{Spatial stability}

We now seek spatial perturbations around a constant profile
that are stationary.
{
They will be given by  solutions for $\Lambda$  of the characteristic equation
det M~$=0$
for $\omega=0$ (stationary solutions).
It is of no surprise that one solution is $\Lambda=1$,
as was already found in the previous subsection.
But here we are not interested in this solution,
which leads to uniform perturbations.
}

{
In order to find the other solutions,
one needs to solve a cubic equation
}
\begin{align}\label{eq:charact_stat}
0=& \ \frac{r}{p}\left[ 1+2\rho^* -4{\rho^*}^2 - \Lambda^{-1}\left( 1-2\rho^* + 8{\rho^*}^2 \right) \right]\\ 
& -(\Lambda-1)\left[  \rho^*(1-4{\rho^*}^2) - \Lambda^{-1}\left( 1-2\rho	^* -4{\rho^*}^2  +8{\rho^*}^3\right) + 4{\rho^*}^2\Lambda^{-2}(1-2\rho^*) \right] \nonumber.
\end{align} 
{
The solutions obviously depend on the bulk plateau density $\rhostar$.
We find that for a low density $\rhostar$, the three roots are real. 
Interestingly, above a critical density $\tilde{\rho}_c$ two of these real
roots
become a complex conjugate pair.
}


For the plateau density of Figure~\ref{logplot},
namely {$\rhostar\simeq0.49918> \tilde{\rho}_c$} we find that the real root $\Lambda_1$
has a modulus smaller than $1$. This corresponds to perturbations
that are exponentially damped for increasing lattice index.
Such a perturbation is indeed observed on the numerical density profiles
of Figure~\ref{logplot}, near the entrance of the system.
The damping rate predicted by the linear stability analysis (slope
of the straight line
in Figure~\ref{logplot}~(b)) is in perfect agreement with the numerics.

The two other complex conjugate roots $\Lambda_2$ and $\Lambda_3$
correspond to oscillatory density perturbations.
We find that above a density $\tilde{\rho}_m$ such that 
$ \tilde{\rho}_c < \tilde{\rho}_m < \rho_0$,
the modulus of $\Lambda_2$ and $\Lambda_3$ is larger than $1$.
This explains that the oscillatory perturbations observed
for $\rhostar=\rho_0$ on the right of the dense region
(see Figure~\ref{logplot}~(a))
are damped when the lattice index \textit{decreases}.
The modulus gives the damping of the oscillatory perturbation 
when it penetrates into the bulk of the dense region.
Although  not perfect, 
there is still a good agreement with the damping observed in the numerics (see
Figure~\ref{logplot}~(b)).

Besides,
the wavelength of the oscillations predicted by the linear
analysis for the parameters of Figure~\ref{logplot}
is $6.85$, in good agreement with the numerical result
of Figure~\ref{logplot}~(a).


In conclusion, although $\rho_0$ still has to be determined
from the numerics, the linear analysis explains { to a large extent
the structure of the density profiles in the dense region.
Again, however,   the shock is moving
so the assumption of stationarity of profiles is not fully accurate.
Instead there is  non-stationarity orginating from the motion of the shock between the discrete lattice sites.
It would be of interest to understand better  how  the propagation of spatial oscillations in the profile and the  dynamics of the shock front  are coupled.
}


\section{Transition between the low density phase and the pulsing phase.}

\begin{figure}[htb]
\centering
\includegraphics[width = 0.7\textwidth]{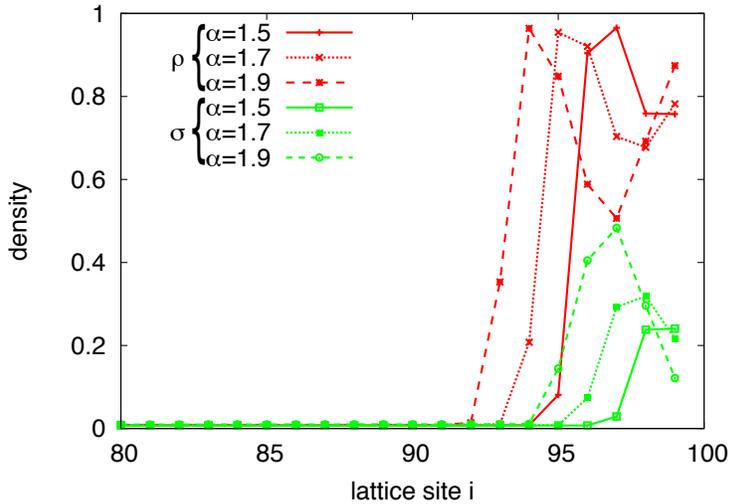}
\caption{The two densities at the reflecting boundary close to the phase transition. We show $\rho$ and $\sigma$ for different values
of the injection rate $\alpha$ as the discontinuous transition is approached at $\alpha \approx 1.9$. \\
Parameter: $r=0.5$, $p=200$, $L=100$.}
\label{transition}
\end{figure}

In this section we try to understand how the transition 
between the pulsing and low density phase is characterised.
We saw in Figure~\ref{phase_transition} that the transition is
discontinuous with respect to an order parameter defined by the
average density through the system.
In Figure \ref{transition} we show the density profiles
for values of $\alpha$ close to the
phase transition. One observes that with increasing $\alpha$ the single peak in $\rho$ close to the
reflecting boundary detaches from the boundary and at the same time
a peak in $\sigma$ arises.
{
Once both densities approach the value one half
at a certain site of the lattice, the transition from the low density phase to
the pulsing phase occurs.
}

We can understand the phase transition better from the particle picture. In the low density phase a boundary
layer of plus particles exists at the right {reflecting boundary}. After some time these plus particles transmute and 
can then leave the system again. If $\alpha$ is low these minus
particles have enough time to leave the dense region at the reflecting boundary. 
{On} increasing $\alpha$, more and more plus particles arrive before
those in the boundary layer transmute
{into minus particles} and leave. In that case those minus particles close to the boundary cannot 
leave the system anymore. The density in both, plus and minus particles, increases and approaches 1/2.
This mutual blocking is necessary for the system to fill up and then trigger the pulsing behaviour. 

\section{Conclusion}

In this work we {have}  presented two models which exhibit a spontaneous pulsing behaviour.
At first we {considered} a particle model inspired by~\cite{lin_s_a2011} where two types of particles
\bluew{can} hop along two lattice lanes of length $L$. At the left boundary, plus particles
are injected and minus particles \bluew{can} leave while the right boundary is a reflecting one.
The two types of particles \bluew{($+$ and $-$)} differ in their hopping direction but also in
their lane changing behaviour. Plus particles cannot change lanes whereas  minus particles can
if the next site in the same lane is already occupied by another particle. Along the
whole system particles can {transmute and} convert their type.
For $p$ large enough so that particles invade the system,
the phase diagram is divided into two phases,
a low density phase and a pulsing phase where 
the system shows spontaneous pulsing in the sense that it fills up
until the density is almost one and then empties again.

Inspired by this explicit particle model we introduced a discrete MF model.
It is defined by the master equations (\ref{bulk_2l_p1}--\ref{right_2l_2}) in the particle densities $\rho$ (for plus particles)
and $\sigma$ (for minus particles). We find that it also exhibits either a very
regular pulsing behaviour or a low density phase (excluding the trivial
case where particles do not change their type).
We start our analysis in this low density phase, for which we can predict the stationary density profile.

{The pulsing phase exhibits a more complex spatial 
structure and temporal behaviour, which makes  the analysis more involved.}
In the filling stage
we can predict analytically the filling
velocity for a small transmutation rate.
In the emptying stage the system's dynamics is very slow.
A front moves {to the left}, separating a dense region on the left
and a dilute region on the right.
In the dense region (between the entrance
and the shock position), both densities deviate from $\frac{1}{2}$ only by very small values, typically $10^{-3}$ or smaller.
In a first approximation we neglect transmutation and predict the shock's
velocity as a function of this deviation, by a mass
conservation argument.
The agreement with the numerics is very good (Figure \ref{velof}).
In the dense region, the density for both types of particles is mostly constant in the bulk.
However, some deviations from the plateau densities are observed
at both ends of the dense region.
Near the entrance, these deviations  decay purely exponentially,
while near the shock the perturbations additionally exhibit some oscillations in the dense region.

{
A linear analysis in the dense region allows {us} to predict
the nature of the perturbations, the damping rate for
each of them, and the wavelength of the oscillatory ones,
with a good agreement with the numerics.
}


In the dilute region, one observes that some {microscopic} density pulses are released
from the shock and propagate through the system to the right boundary.
This {microscopic} pulse emission is connected to the hopping of the shock from one lattice
site to the next one.

Eventually, the transition from the low density phase to the pulsing phase
appears to occur through the detachment of an accumulation of particles from the reflecting  boundary. {The resulting transition is discontinuous.}



{
For this system, several challenging questions are still open.
Though we were able to characterise the low density phase,
and to explain the structure of density profiles in the pulsing phase,
in the emptying stage of the pulsing phase
we are still lacking a full prediction of the shock velocity---and
hence of the pulsing period.
The fact that each particle type density is not conserved
locally adds a further complication.
It would be of interest  to be able to take it into account
in some generalised mass conservation relations.
The presence of steep gradients separating constant regions
makes it difficult to use perturbative calculations.

{We have also seen how  the   motion of the shock  between
the discrete  lattice sites gives  rise to microscopic density pulse emissions which in turn imply non-stationary profiles.
There are many  systems, for example   involving  molecular motor dynamics on filaments, where 
discreteness of a spatial  lattice is imposed.
Therefore  would be of interest to further explore 
the mechanisms  of such {microscopic} pulses.}
}

\ack
We thank Danielle Hilhorst, Fr\'ed\'eric Lagouti\`ere, Christian Tenaud,
Henk Hilhorst and Pascal Viot for
fruitful discussions.

This work was supported by ``Investissements d’Avenir'' of LabEx
PALM (ANR-10-LABX-0039-PALM) --
Funding agency : Investissement d'Avenir LabEx PALM --
Grant number : ANR-10-LABX-0039

S.K. was supported by the Deutsche 
Forschungsgemeinschaft (DFG) within the collaborative 
research center SFB 1027 and the research training group GRK 1276.

M.R.E. acknowledges partial support from EPSRC
under grant number EP/J007404/1 {and thanks LPT for hospitality.}

\vskip 2em\appendix{}

\section{Order by order expansion in the low density phase}
The lowest order of the expansion~(\ref{taylor_exp_K})  yields
\begin{equation}
\rho^{(1)} - \sigma^{(1)} = 0
+ \mathcal{O}\left(\frac{1}{\LL^3}\right)
\end{equation}
while the lowest order of Eqs~(\ref{taylor_exp_rho},\ref{taylor_exp_sig})
gives (in the stationary state)
\begin{eqnarray}
0 &=& - r \left( \rho^{(1)} - \sigma^{(1)}\right)
   - \frac{p}{\LL}
   \frac{\partial \rho^{(1)}}{\partial x}
   + \mathcal{O}\left(\frac{1}{\LL^3}\right) 
\label{dtr_order1} \\
0 &=& r \left( \rho^{(1)} - \sigma^{(1)}\right)
   + \frac{p}{\LL}
   \frac{\partial \sigma^{(1)}}{\partial x}
   + \mathcal{O}\left(\frac{1}{\LL^3}\right)
\label{dts_order1}
\end{eqnarray}
As a result, one finds that $\frac{\partial \sigma^{(1)}}{\partial x} = 0$ and
$\frac{\partial \rho^{(1)}}{\partial x} = 0$.
Then we get the constant value of $\rho^{(1)} = \sigma^{(1)}$ from the entrance
boundary
\begin{equation}
\rho^{(1)}(x) = \sigma^{(1)}(x) = \sigma^{(1)}(0) = \frac{\alpha}{p}
\end{equation}

We can show that the next order brings no new term, so that
\begin{equation}
\rho^{(2)}(x) = \sigma^{(2)}(x) = 0.
\end{equation}
We are thus left with
\begin{eqnarray}
\rho & = & \frac{\alpha}{p} + \rho^{(3)} + \mathcal{O}\left(\frac{1}{\LL^5}\right)\\
\sigma & = & \frac{\alpha}{p} + \sigma^{(3)} + \mathcal{O}\left(\frac{1}{\LL^5}\right)
\end{eqnarray}

To next order,
assuming that $\frac{\partial \rho^{(3)}}{\partial x} = \mathcal{O}\left(\frac{1}{\LL^4}\right)$ and $\frac{\partial \sigma^{(3)}}{\partial x} = \mathcal{O}\left(\frac{1}{\LL^4}\right)$,
the no flux condition  
(\ref{taylor_exp_K})  
gives
\begin{equation}
0 = -2 \left(\frac{\alpha}{p}\right)^2
+ \rho^{(3)}(x) - \sigma^{(3)}(x)
+ \mathcal{O}\left(\frac{1}{\LL^5}\right)
\label{K0_order3}
\end{equation}
and the bulk equations yield
\begin{eqnarray}
0 &=& - r \left( \rho^{(3)} - \sigma^{(3)}\right)
   - \frac{p}{\LL}
   \frac{\partial \rho^{(3)}}{\partial x}
   + \mathcal{O}\left(\frac{1}{\LL^5}\right)
\label{dtr_order3} \\
0 &=& r \left( \rho^{(3)} - \sigma^{(3)}\right)
   + \frac{p}{\LL}
   \frac{\partial \sigma^{(3)}}{\partial x}
   + \mathcal{O}\left(\frac{1}{\LL^5}\right)
\label{dts_order3}
\end{eqnarray}
The sum of these last equations gives
\begin{equation}
\frac{\partial \rho^{(3)}}{\partial x}
= \frac{\partial \sigma^{(3)}}{\partial x}
= - 2 \frac{\LL}{p} r \left(\frac{\alpha}{p}\right)^2
\end{equation}
where the last equality is obtained by using (\ref{K0_order3})
into (\ref{dts_order3}).

Now combining (\ref{s1_ness}) and (\ref{K0_order3}) taken for $x=0$,
we get
\begin{equation}
\rho^{(3)}(0) = 0 \;\;\; \mbox{ and } \;\;\;
\sigma^{(3)}(0) = -2 \left(\frac{\alpha}{p}\right)^2.
\end{equation}

Collecting all terms, we have eventually
\begin{eqnarray}
\rho & = & \frac{\alpha}{p} - \frac{2 r \LL}{p} \left(\frac{\alpha}{p}\right)^2 x
+ \mathcal{O}\left(\frac{1}{\LL^5}\right)
\label{rho_lowdens2} \\
\sigma & = & \frac{\alpha}{p}\left[1 - 2 \frac{\alpha}{p}\right] - \frac{2 r \LL}{p} \left(\frac{\alpha}{p}\right)^2 x
+ \mathcal{O}\left(\frac{1}{\LL^5}\right)
\label{sigma_lowdens2} 
\end{eqnarray}






\clearpage

\end{document}